\documentclass{sig-alternate-2013}

\newfont{\mycrnotice}{ptmr8t at 7pt}
\newfont{\myconfname}{ptmri8t at 7pt}
\let\crnotice\mycrnotice%
\let\confname\myconfname%

\usepackage[
  pass,
]{geometry}

\permission{\rev{Permission to make digital or hard copies of all or part of this work for personal or classroom use is granted without fee provided that copies are not made or distributed for profit or commercial advantage and that copies bear this notice and the full citation on the first page. Copyrights for components of this work owned by others than ACM must be honored. Abstracting with credit is permitted. To copy otherwise, or republish, to post on servers or to redistribute to lists, requires prior specific permission and/or a fee. Request permissions from permissions@acm.org.}
\conferenceinfo{\rev{MobiHoc'15,}}{\rev{June 22--25, 2015, Hangzhou, China.}}
\copyrightetc{\rev{Copyright \copyright~2015 ACM \the\acmcopyr}}
\crdata{978-1-4503-3489-1/15/06\ ...\$15.00.\\
http://dx.doi.org/10.1145/2746285.2746293}}

\ifx\UseOption\undefined
\def\UseOption{opt2} 
\fi

\makeatletter
\def\@copyrightspace{\relax}
\makeatother

\ifodd 0
\newcommand{\rev}[1]{{\color{blue}#1}}
\else
\newcommand{\rev}[1]{#1}
\fi

\ifodd 0

\else

\fi

\def\QEDclosed{\mbox{\rule[0pt]{1.3ex}{1.3ex}}} 

\usepackage{amssymb,amsmath,color,graphicx}
\usepackage{subfigure}
\usepackage{cite}
\usepackage{verbatim}
\usepackage{algorithm}
\usepackage{algorithmic}
\usepackage{url}
\usepackage{graphicx,epstopdf}
\usepackage{optional}

\DeclareMathOperator*{\argmax}{arg\,max}

\newtheorem{theorem}{Theorem}

\newtheorem{definition}{Definition}

\newtheorem{lemma}{Lemma}

\begin{document}

\title{Distributed Time-Sensitive Task Selection \\ in Mobile Crowdsensing}

\numberofauthors{1} 
\author{\alignauthor \rev{Man Hon Cheung\textsuperscript{*\#}, Richard Southwell\textsuperscript{\#}, Fen Hou\textsuperscript{*}, and Jianwei Huang\textsuperscript{\#}} \\
\affaddr{\rev{\textsuperscript{*}Department of Electrical and Computer Engineering, University of Macau, Macau}} \\
\affaddr{\rev{\textsuperscript{\#}Department of Information Engineering, The Chinese University of Hong Kong, Hong Kong, China}} \\
\email{\rev{mhcheung@ie.cuhk.edu.hk, richardsouthwell254@gmail.com, fenhou@umac.mo, jwhuang@ie.cuhk.edu.hk}}
}

\maketitle


\begin{abstract}
  With the rich set of embedded sensors installed in smartphones and the large number of mobile users, we witness the emergence of many innovative commercial mobile crowdsensing applications that combine the power of mobile technology with crowdsourcing to deliver time-sensitive and location-dependent information to their customers.
  Motivated by these real-world applications, we consider the task selection problem for heterogeneous users with different initial locations, movement costs, movement speeds, and reputation levels.
	Computing the social surplus maximization task allocation turns out to be an NP-hard problem.
  Hence we focus on the distributed case, and propose an asynchronous and distributed task selection (ADTS) algorithm to help the users plan their task selections on their own. We prove the convergence of the algorithm, and further characterize the computation time for users' updates in the algorithm.
  Simulation results suggest that the ADTS scheme achieves the highest Jain's fairness index and coverage comparing with several benchmark algorithms, while yielding similar user payoff to a greedy centralized benchmark. Finally, we illustrate how mobile users coordinate under the ADTS scheme based on some practical movement time data derived from Google Maps.
\end{abstract}

\category{\rev{C.2.4}}{\rev{Computer-Communication Networks}}{\rev{Distributed Systems}}[\rev{Distributed applications}]


\keywords{\rev{Mobile crowdsensing; crowdsourcing; game theory}} 

\section{Introduction} \label{sec:intro}

  Today's smartphones and wearable devices include a rich set of \emph{embedded sensors}, such as cameras, microphones, global positioning systems (GPS), thermometers, and accelerometers \cite{lane_as10, ganti_mc11}. Thanks to the large number of mobile users and their inherent mobility, we are witnessing the rise of \emph{mobile crowdsensing} (MCS), where individuals use their mobile devices to collectively extract and share information related to some phenomenon of interest. 
  Applications of MCS include traffic jam alerts, wireless indoor localization, and small cell network monitoring. 

  Recently, commercial MCS platforms, such as Gigwalk \cite{gigwalk} and Field Agent \cite{fieldagent}, combine the power of mobile technology with crowdsourcing to provide their customers with \emph{time-sensitive} and \emph{location-dependent} information related to their stores and products.
  For example, in Field Agent \cite{fieldagent}, there are two types of \emph{tasks}, namely audit and research. An audit task mainly involves fact finding and data gathering (such as checking on-shelf availability and prices), while a research task mainly involves the collection of users' opinions and insights (such as surveys and shop-alongs).
  For a location-dependent and time-sensitive sensing task, such as checking the on-shelf availability of Coca-Cola in a convenience store on Main Street at 9am, the users need to collect the data at the \emph{precise} time and location. With this information, a district manager can reduce the cost of taking inventories, while maintaining the proper stock levels at different stores.
  Currently, several well-known brands and retailers (such as Coca-Cola and Johnson \& Johnson) are among the customers of Gigwalk and Field Agent. This suggests that the collection of location-dependent and time-sensitive data using MCS is a practice of growing importance.

  A key question for MCS platforms is \emph{how to find users to complete tasks, while accounting for the users' different initial locations, movement costs, movement speeds, and reputation levels?} A number of recent results have focused on the \emph{centralized} task allocation in MCS, with the objectives of either improving the energy efficiency or maximizing the social surplus.
  For example, Sheng \emph{et al.} in \cite{sheng_ee12} considered opportunistic energy-efficient collaborative sensing. Given a set of roads, mobile devices, and their trajectories, the objective is to find a sensing schedule that minimizes the total energy consumption by reducing redundancy in sensing.
  Zhao \emph{et al.} in \cite{zhao_fe14} considered fair and energy-efficient task allocation in MCS, and solved a min-max aggregate sensing time problem. 
   He \emph{et al.} in \cite{he_to14} considered social surplus maximization for location-dependent task scheduling in MCS. They formulated a task scheduling problem, where the objective is to maximize the overall net reward of all the users, subject to the time budget of each user and the sensing redundancy constraint of each sensing task. 

  Different from the above literature, which focuses on the collection of location-dependent data \emph{without} any time constraints in a \emph{centralized} fashion, we consider the case where the service provider aims to collect \emph{time-sensitive} and location-dependent information for its customers through \emph{distributed} decisions of mobile users. The solution of such a problem needs to balance the rewards and movement costs of the users for completing tasks. The consideration along both space and time dimensions makes the design of a good solution very challenging. The distributed nature of the solution that is required by the current commercial platform such as Gigwalk and Field Agent further complicates the analysis.

 In this paper, we focus on solving the distributed time-sensitive and location-dependent task selection problem, where users are heterogeneous in their initial locations, movement costs, movement speeds, and reputation levels.
  We formulate the interactions among users as a non-cooperative task selection game (TSG), and propose an asynchronous and distributed task selection (ADTS) algorithm for each user to compute her task selection and mobility plan. 
  Each user only requires limited information on the aggregate task choices of all users, which is publicly available in many of today's crowdsourcing platforms, such as Amazon Mechanical Turk (MTurk) \cite{mturk}, Gigwalk \cite{gigwalk}, and Field Agent \cite{fieldagent}.
	
  As a \emph{performance benchmark} on the social surplus, we formulate a centralized task allocation (CTA) problem, assuming that all users are controlled by the service provider, and show that it is an NP-hard problem. We propose a heuristic greedy centralized algorithm that computes the approximate centralized solution with a lower complexity.

  To the best of our knowledge, this is the first paper that considers distributed time-sensitive and location-dependent data collection in MCS motivated by real-life commercial applications.
  To summarize, the contributions of our work are as follows:
\begin{itemize}

\item \emph{Practical MCS modeling}: Motivated by commercial MCS applications, we consider the collection of time-sensitive and location-dependent sensing data by multiple users. We assume that the users are heterogeneous in their initial locations, movement costs, movement speeds, and reputation levels. 

\item \emph{Asynchronous and distributed task selection algorithm}: We propose a distributed  algorithm that helps the users determine their task selections and mobility plans. Each user only needs to know limited information on the aggregate task choices readily available in the MCS mobile apps.

\item \emph{Convergence guaranteed}: We show that the task selection game has the finite improvement property, which means that users' asynchronous best response updates globally converge to a Nash equilibrium. We also show that each best response update can be computed in polynomial time.

\item \emph{Balanced performance}: Simulation results suggest that the proposed asynchronous and distributed task selection scheme achieves the best performance in terms of fairness and coverage comparing with various benchmark algorithms.


\end{itemize}

\section{System Model} \label{sec:model}


\begin{figure}[t]
 \centering
   \includegraphics[width=7cm]{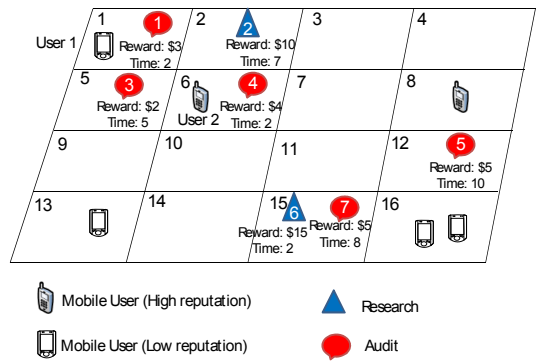} 
 \caption{In a MCS platform, we assume that the service provider collects location-dependent time-sensitive information with the help of the users. Through the mobile app interface, the service provider would announce the rewards, locations, and execution times of the tasks to the users. Each user then decides how to move and which tasks to work on over the coming time slots.} 
\label{fig:map}
\end{figure}

  We consider a mobile crowdsensing (MCS) platform that involves the collection of \emph{time-sensitive} and \emph{location-dependent} information in the form of photos, video, audio, data, opinions, and feedback.
  In the platform, once a user has installed the mobile app on her smartphone, she will be able to use the built-in map to check the available tasks, together with the task attributes including locations, execution times, and rewards, as shown in Fig.~\ref{fig:map}.

  More specifically, let $\mathcal{L}=\left\{ 1,\ldots,L\right\}$ be the set of \emph{locations}, such as buildings, landmarks, or fields. Let $\mathcal{T}=\left\{ 1,\ldots,T\right\}$ be the set of \emph{time slots}. Let $\mathcal{K}=\left\{ 1,\ldots,K\right\}$ be the set of \emph{tasks}. We describe the attributes related to a task as follows.


\begin{definition}
  \normalfont{(Task Attributes) }
Each task $k\in\mathcal{K}$ is associated with:
\begin{itemize}
\item A location $l[k]\in \mathcal{L}$ \emph{where} the task must be performed.

\item A time $t[k]\in\mathcal{T}$ \emph{when} the task must be executed\opt{opt2}{\footnote{Each task $k$ only takes one time slot to execute, and it cannot be executed before the corresponding time $t[k]$.}}.

\item A \emph{reward} amount $\rho[k]\geq0$ for performing the task. In the case where multiple users perform the same task, the reward is shared evenly among them \cite{duan_ms14}. 

\end{itemize}
\end{definition}

  For example, for task $5$ in Fig.~\ref{fig:map}, we have $l[5] = 12$, $t[5] = 10$, and $\rho[5] = \$5$. Notice that there can be multiple tasks at the same location, such as $l[6] = l[7] = 15$.

  Let $\mathcal{I} = \{1,\ldots,I\}$ be the set of users, who can move through space and perform tasks. Given a user's initial location, the set of tasks that she chooses to work on is influenced by her \emph{movement time}, \emph{movement cost}, and \emph{reputation}.
  Formally, we define a user's attributes as follows.

\begin{definition}
  \normalfont{(User Attributes) }
Each user $i\in\mathcal{I}$ is associated with:
\begin{itemize}
\item An \emph{initial location} $l_i ^{\text{init}} \in \mathcal{L}$. (For example, we have $l_2^{\text{init}} = 6$ in Fig.~\ref{fig:map}.)

\item An (integer valued) \emph{movement time} $\Delta_{i}^{l,l'}\geq1$, which equals the number of time slots it takes user $i$ to move from location $l$ to location $l'.$

    \item A (real valued) \emph{movement cost} $c_{i}^{l,l'}\geq0$ for user $i$ to move from
location $l$ to location $l'.$

\item A non-empty set $\mathcal{K}_{i}\subseteq\mathcal{K}$ of tasks that user $i$ is eligible for.
\end{itemize}
\end{definition}

  If a user has specified her mode of transportation (e.g., walking, cycling, driving, or taking the bus), then her mobile device can retrieve data from Google Maps and local public transportation databases to compute her movement time and movement cost.
  For example, user $i$ may have a zero cost (i.e., $c_i^{{l,l'}} = 0$) for walking, and a linear cost for driving\opt{opt2}{\footnote{We can set the movement cost $c_i^{l,l'} = \infty$ if a user $i$ is unwilling or unable to travel from location $l$ to location $l'$.}}. 

  The set of eligible tasks $\mathcal{K}_{i}$ often relates to the \emph{reputation} of user $i$. Reputation mechanisms are commonly found in crowdsensing applications, such as Gigwalk \cite{gigwalk} and MTurk \cite{mturk}, to allocate tasks based on users' past performance. A service provider can reserve the more complicated but higher paying tasks (e.g., research tasks) for users with higher reputations. For example, in Fig.~\ref{fig:map}, the lower reputation user $1$ can only work on the audit tasks, so $\mathcal{K}_{1} = \{1,3,4,5,7\}$; whereas the higher reputation user $2$ can work on both the audit and research tasks, so $\mathcal{K}_2 = \{1,\ldots,7\}$.

   Given the rewards, locations, and execution times of different tasks, each user needs to decide how she should move and which tasks she should select in order to maximize her total profit (i.e., reward minus movement cost).
	In Section \ref{sec:distributed}, we formulate the
task selection of multiple users as a task selection game, and propose an ADTS algorithm. In Section \ref{sec:centralized}, we establish a performance benchmark under the ideal centralized task allocation scenario. We use simulations to evaluate the performance of our proposed algorithms in Section \ref{sec:pe}, and draw conclusions in Section \ref{sec:concl}.

\section{Distributed Task Selection Games} \label{sec:distributed}


  In this section, we formulate the users' task selection problem as a \emph{task selection game (TSG)}, where each user aims to maximize her own payoff. We also study the equilibria and convergence properties of the game. 

\subsection{Task-Time Routing} \label{sec:tasktimerouting}

  A user's decision making includes two aspects. First, she must choose which locations to visit over the $T$ time slots (i.e., selecting a route through space-time). Second, she must choose which tasks to do along that route. However, since each task $k$ is only identified with a single location $l[k]$ and time $t[k]$, we can simplify the modeling by just considering how the users select different tasks at different times (i.e., choosing a route through \emph{task-time}).

  Since a user may choose not to work on any task, for each user $i$ we define an initial virtual task $k_i ^{\text{init}}\in \mathcal{K}_i$, which occurs at all time slots $t[k_i ^{\text{init}}] \in \mathcal{T}$ and at the same location $l[k_i ^{\text{init}}]=l_i ^{\text{init}}$ (user $i$'s initial location) with a zero reward (i.e., $\rho[k_i ^{\text{init}}]=0$).

  To clearly describe the task-time routing decision problem, we use a sequences of task-time points, described in Definition \ref{def:tasktimeroute}, to represent a user's task-time route. 
  We say that a task-time route is available to a user when it is physically possible for her to visit the corresponding locations over the time slots, and she is eligible to perform each of the tasks at those locations.

\begin{definition} \label{def:tasktimeroute} 
  \normalfont{(Available Task-Time Route) }
An \emph{available task-time route for a user $i$} is a sequence
\begin{equation} \label{equ:ri}
	r_i = \left(\left(k_i^{1},t_i^{1}\right),\left(k_i^{2},t_i^{2}\right),\ldots,\left(k_i^{n},t_i^{n}\right)\right)\in \left(\mathcal{K}\times \mathcal{T}\right)^n
\end{equation}
of $n$ task-time points (for some $n\geq1$) which satisfies the following conditions:
 \begin{enumerate}
  \item \emph{Time increases:} $1=t_i^{1}<t_i^{2}<\ldots<t_i^{n} \leq T.$
 \item \emph{The user is eligible for the tasks:} $k_i^{1},k_i^{2},\ldots,k_i^{n}\in \mathcal{K}_i.$
 \item \emph{Sequence starts at the initial virtual task}: $k_i^1 = k_i^{\text{init}}.$
 \item\emph{Sequence accounts for movement time:} $t_i^{a+1}-t_i^a=\Delta_i^{l[k_i^a],l[k_i^{a+1}]}$, for each $a\in \{1,\ldots,n-1\}$.
 \end{enumerate}
\end{definition}

Condition $1$ accounts for the fact that time is always increasing. Condition $2$ ensures that the users are eligible to perform each of their chosen tasks. Condition $3$ ensures that the user begins at her initial location. Condition $4$ ensures that the time difference between successive elements in the sequence of task-time elements is equal to the movement time between the locations of the corresponding tasks. (In other words, it ensures that the movement time is indeed the time they spend moving.) 
  When a user does not move, we define $\Delta_i^{l,l} = 1, \, \forall \, i \in \mathcal{I}, l \in \mathcal{L}$ to represent the fact that the user stays at the same location after one time slot.

\begin{figure}[t]
 \centering
   \includegraphics[width=6.5cm, height=4.5cm]{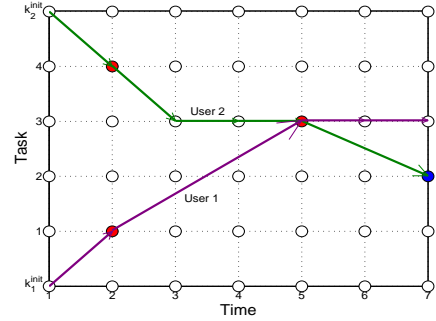} 
 \caption{Task-time routes chosen by two users with audit (red circles) and research (blue circle) tasks.} 
\label{fig:route}
\end{figure}

  Based on the example in Fig.~\ref{fig:map} by considering task $1$ to task $4$, we show an example of the task-time routes of two users in Fig.~\ref{fig:route}. Here the solid and empty circles represent task-time points with and without positive rewards, respectively. When a user's route passes through a solid circle, it means that the user works on the task (e.g., user $2$ works on task $3$ at time $5$). On the other hand, when a user's route passes through an empty circle, it means that the user is only physically present at the location of a task, but she is not working on the task (e.g., user $2$ is at location $l[3]$ at time $3$). In this way, a user can move to the location of a task before the actual execution time of that task (e.g., user $2$ arrives at location of task $3$ at time $3$, while the execution time of task $3$ is at time $5$).
	
	In Fig.~\ref{fig:route}, we suppose that a low reputation user $1$ has movement times $\Delta_1 ^{l_1 ^{\text{init}},l[1]} = 1$ and $\Delta_1 ^{l[1],l[3]} = 3$, while a high reputation user $2$ has movement times $\Delta_2 ^{l_2 ^{\text{init}},l[4]} =1$, $\Delta_2 ^{l[4],l[3]} = 1$, and $\Delta_2 ^{l[3],l[2]} = 2$. 
	The purple and green curves represent the two task-time routes: $r_1 = \bigl((k_1^{\text{init}}, 1), (1,2), (3,5),$ $(3,6), (3,7) \bigr)$ and $r_2 = \bigl((k_2^{\text{init}}, 1), (4,2), (3,3), (3,4), (3,5), (2,7) \bigr)$.



\subsection{Task Selection Game}

Based on the discussion aforementioned, we can formulate the users' task selection problem as a TSG, where
users act as players that choose available task-time routes. 

  In the task-time routing framework, changing tasks will often involve changing locations and hence will involve movement costs. To allow a user to move to the location of a task before its execution time, we define the \emph{time-dependent reward} $\rho^{*}[k,t']$ for task $k\in \mathcal{K}$ and time $t' \in \mathcal{T}$ as
$$\rho^{*}[k,t']=\begin{cases}
\rho[k], & \mbox{ if }t'=t[k],\\
0, & \mbox{ otherwise.}
\end{cases}$$
  For example, in Fig.~\ref{fig:route}, the rewards for task-time points $(3,3)$ and $(3,5)$ are $\rho^*[3,3]=0$ and $\rho^*[3,5]=\rho[3]=\$2$, respectively.
  We assume that the reward $\rho^{*}[k,t']$ is evenly shared among all the users who have chosen to work on task $k$ at time $t'$ (i.e., the task-time point $(k,t')$.)

\begin{definition}
  \normalfont{(Task-time points)}
 Let us define the \emph{task-time points} of a task-time route $r_i$ in \eqref{equ:ri} to be the set $\mathbb{V}(r_i) = \{ \left(k_i^{1},t_i^{1}\right),\left(k_i^{2},t_i^{2}\right),\ldots,\left(k_i^{n},t_i^{n}\right)\},$ of all task-time points visited by the route $r_i \in \mathcal{R}_i$, where $\mathcal{R}_i$ denotes the set of all available task-time routes of user $i$.
\label{z:tt points} 
\end{definition}

The \emph{payoff} $U_i(\boldsymbol{r})$ that user $i$ gets for choosing route $r_i$ in \eqref{equ:ri} in a strategy profile $\boldsymbol{r}=(r_1,\ldots,r_I)\in \mathcal{R}_1 \times \ldots \times \mathcal{R}_I$ is equal to
\begin{equation} \label{equ:payoff}
U_i(\boldsymbol{r})=\left(\sum_{a=1}^n \frac{\rho^* [k_i^a,t_i^a]}{m^{(k_i^a,t_i^a)}(\boldsymbol{r})}\right) -  \sum_{a=1}^{n-1} c_i ^{l[k_i^a],l[k_i^{a+1}]},
\end{equation}
where $m^{(k,t)}(\boldsymbol{r})=|\{j\in \mathcal{I}:(k,t) \in \mathbb{V}(r_j)\}|$ is the number of routes that pass through task-time point $(k,t)$. 
  The first term in \eqref{equ:payoff} corresponds to the total reward that user $i$ obtains (taking into account how the reward is shared evenly when multiple users perform the same task). The second term in \eqref{equ:payoff} corresponds to the total movement cost, which user $i$ spends in order to travel to the locations of the selected tasks.
%
\begin{definition}
  \normalfont{(Task selection game) }
A \emph{task selection game}
$	\Omega = \bigl(\mathcal{I},\mathcal{L},\mathcal{K},\mathcal{T},(\rho^*[k,t])_{k\in \mathcal{K},t\in \mathcal{T}},
	(c_i^{l,l'},\Delta_i^{l,l'})_{i \in \mathcal{I}, l,l'\in \mathcal{L}}\bigr),$
involves each user (player) $i$ choosing an available task-time route (strategy) $r_i \in \mathcal{R}_i$ and receiving the payoff in \eqref{equ:payoff}.
\end{definition}

\begin{definition}
  \normalfont{(Task-time point pairs)}
 The \emph{task-time point pairs $\mathbb{E}(r_i)$} of a task-time route $r_i$ in \eqref{equ:ri} is the set $\mathbb{E}(r_i) = \Bigl\{ \left[ \left(k_i^{a},t_i^{a}\right),\left(k_i^{a+1},t_i^{a+1}\right) \right]: a = 1, \ldots, n-1 \Bigr\}$ of task-time point pairs subsequently visited by the route $r_i$.
 \label{z:ttp points}
\end{definition}

  Let $g_i\Bigl( \bigl[\left(k_i^{a},t_i^{a}\right),\left(k_i^{a+1},t_i^{a+1}\right)\bigr] \Bigr) = c_i^{l[k_i^a],l[k_i^{a+1}]}$ be the movement cost between two adjacent task-time points of user $i$.
  The payoff of user $i$ in \eqref{equ:payoff} can be rewritten as
\begin{equation} \label{equ:payoff2}
U_i(\boldsymbol{r})=\left(\sum_{(k,t) \in \mathbb{V}(r_i)} \frac{\rho^* [k,t]}{m^{(k,t)}(\boldsymbol{r})}\right) -  \sum_{e \in \mathbb{E}(r_i)} g_i(e).
\end{equation}

\subsection{Equilibrium Existence and Convergence Analysis}

  Before analyzing the task selection games, let us recall some commonly used definitions from game theory. Let $\boldsymbol{r}_{-i} = (r_1, \ldots, r_{i-1}, r_{i+1}, \ldots, r_I)$ denote the strategies of all the users except user $i$. A strategy profile can be written as $\boldsymbol{r} = (r_i, \boldsymbol{r}_{-i}).$
\vspace{0.2cm}
\begin{definition}
  \normalfont{(Better and best response updates) }
   A \emph{better response update}, starting from some strategy $\boldsymbol{r} = (r_i, \boldsymbol{r}_{-i})$, is an event where a single user $i$ changes to another strategy, $r_i' \in \mathcal{R}_i$, and increases her payoff as a result, i.e., $U_i(r_i', \boldsymbol{r}_{-i}) > U_i(r_i, \boldsymbol{r}_{-i})$.

   A \emph{best response update} is a special type of better response update, where the newly selected strategy $r_i'$ maximizes user $i$'s payoff among user $i$'s all possible better response updates. 
\end{definition}
\vspace{0.2cm}
\begin{definition} \label{def:NE} 
  \normalfont{(Pure Nash equilibrium) }
A \emph{pure Nash equilibrium (NE)} is a strategy profile $\boldsymbol{r}^*$, where no user can perform a better response update unilaterally.
\end{definition}

\vspace{0.2cm}
\begin{definition} \label{def:FIP} 
  \normalfont{(Finite improvement property) }
   A game possesses the \emph{finite improvement property} (FIP) when asynchronous\opt{opt2}{\footnote{Asynchronous updates imply that there will be no two users updating their strategies at the same time.}} better response updates always converge to a pure NE within a finite number of steps, irrespective of the initial strategy profile or the users' updating order.
\end{definition}

Existence of the finite improvement property implies that better response updating always leads to pure Nash equilibria, which implies the existence of pure Nash equilibria.\opt{opt2}{\footnote{Note that the time steps involved in understanding the convergence of the best response updates is not the same time slot that we introduced in Section \ref{sec:model}.}}

\vspace{0.2cm}
\begin{theorem} \label{THM:FIP}
Every task selection game possesses the finite improvement property.
\end{theorem}
\vspace{0.2cm}

  \opt{opt1}{A proof sketch of Theorem \ref{THM:FIP} is given in Appendix \ref{app:fip}.}
	\opt{opt2}{The proof of Theorem \ref{THM:FIP} is given in Appendix \ref{app:fip}.}
  We then proceed to study how long it takes for a strategy profile to converge to a pure NE.
	Theorem \ref{THM:BRPOLY} ensures that each best response update can be computed in polynomial time. 
\vspace{0.2cm}
\begin{theorem} \label{THM:BRPOLY}
  A best response update can be computed in $\mathcal{O}(K^3 T^3)$ time.
\end{theorem}

  The proof of Theorem \ref{THM:BRPOLY} is given in Appendix \ref{app:brpoly}. 
	The proof is constructive, as it allows us to compute best response updates in polynomial time. Simulations suggest that the number of best response updates required to reach a pure Nash equilibrium grows linearly with the number of users in a wide variety of scenarios, although a theoretical proof is quite challenging to obtain.



\begin{algorithm} [t] \small
\caption{\emph{Asynchronous and distributed task selection (ADTS) algorithm for user $i \in \mathcal{I}$.}}
\begin{algorithmic} [1] \label{algo:tsg_mu}

\STATE \underline{Initialization}

\STATE User Input: Movement cost $c_i^{l,l'} \, \forall \, l,l' \in \mathcal{L}$.

\STATE Check the task description, location $l[k]$, time $t[k]$, and reward $\rho[k]$ for all task $k \in \mathcal{K}$ on the mobile app interface.

\STATE Calculate movement time $\Delta_i^{{l,l'}}, \, \forall \, l, l' \in \mathcal{L}$ using Google Maps data based on movement speed $\nu_i$.



\STATE \underline{Planning Phase: Task Selection Game}

\STATE \textbf{repeat}

\STATE $ \ \ \ $ Check clock timer $\tau$ on the mobile app.

\STATE $ \ \ \ $ \textbf{if} $\tau \in \Gamma_i$

\STATE $ \ \ \ \ \ $ Check the mobile app for the number of participants \\ \quad\quad\quad $q^{(k,t)}$ for all $(k,t) \in \mathcal{H}$.

\STATE $ \ \ \ \ \ $ Calculate the number of participants excluding user $i$ \\ \quad\quad\quad itself:  $q_{-i}^{(k,t)} := q^{(k,t)} - q_i^{(k,t)}, \, \forall \, (k,t) \in \mathcal{H}$.

\STATE $ \ \ \ \ \ $ Perform a best response update: Find a route $r_i \in \mathcal{R}_i$ \\ \quad\quad\quad  that maximizes user $i$'s payoff:  \\ \quad\quad\quad
   $\displaystyle U_i(r_i,\boldsymbol{r}_{-i}) := \left(\sum_{(k,t) \in \mathbb{V}(r_i)} \frac{\rho^* [k,t]}{q_{-i}^{(k,t)}+1}\right) -  \sum_{e \in \mathbb{E}(r_i)} g_i(e)$.

\STATE $ \ \ \ \ \ $ Update the task selection decision:
\begin{equation} 
	  q_i^{(k,t)} :=
\begin{cases}
  1, & \mbox{if } (k,t) \in \mathbb{V}(r_i), \\
  0, & \mbox{otherwise,} 
\end{cases}
\; \forall \, (k,t) \in \mathcal{H}.
\end{equation}

\STATE $ \ \ \ \ \ $ Report $\boldsymbol{q}_i := (q_i^{(k,t)}, \, \forall \, (k,t) \in \mathcal{H})$ to the service provider.

\STATE $ \ \ \ $ \textbf{end if}

\STATE \textbf{until} $\tau \geq \tau^{\text{max}}$.

\STATE \underline{Data Collection and Sensing Phase}

\STATE \textbf{for} each user $i \in \mathcal{I}$

\STATE $ \ \ \ $ Move and complete the sensing task in each time slot $t$   \\ \quad\quad\quad based on the task selection plan $r_i$.

\STATE \textbf{end for}

\end{algorithmic}
\end{algorithm}

%
\subsection{Asynchronous and Distributed Task Selection Algorithm}

  Theorems \ref{THM:FIP} and \ref{THM:BRPOLY} guarantee the convergence of our asynchronous and distributed task selection (ADTS) algorithm (Algorithm \ref{algo:tsg_mu}). 
	To initialize the algorithm, a user inputs her private information on the movement cost (line 2), checks the task descriptions (line 3), and then computes her movement time from Google Maps data\opt{opt2}{\footnote{Google Maps assume an average walking speed of about $3$ miles/hour for pedestrians. For example, if the Google Maps show that movement time between  locations $l$ and $l'$ is $2$ mins, then user $i$ with movement speed of $4$ miles/hour would have a movement time $\Delta_i^{{l,l'}} = 1.5$ mins.}} based on her speed\opt{opt2}{\footnote{Prototype systems, such as BreadCrumbs \cite{nicholson_bf08}, can track the movement speed of the device's owner.}} (line 4).

  With the aggregate information\opt{opt2}{\footnote{Such information is often available for users in crowdsourcing platforms.}} on the total number of users working on different tasks (line 9) provided by the service provider in Algorithm \ref{algo:tsg_sp}, each user performs a best response update (lines 11 and 12), and claims the tasks by sending her updated task selection $\boldsymbol{q}_i = (q_i^{(k,t)}, \, \forall \, (k,t) \in \mathcal{H})$ to the service provider. Here, $\mathcal{H} = \{ (k,t)\in \mathcal{K} \times \mathcal{T}: \rho^*[k,t]>0 \}$ is the set of task-time points that provide a positive reward.

  Notice that users only need to claim the tasks that they are interested in working on, and do not need to reveal their movement plans to the service provider. This will preserve the privacy of users. Let $\Gamma_i$ be the set of iterations\opt{opt2}{\footnote{Note that an iteration only takes a fraction of a single time slot.}} during which user $i$ updates her task selection strategy. We assume that each user updates her strategy distributively and asynchronously until a predefined iteration limit $\tau^{\text{max}}$. 
  We set $\tau^{\text{max}}$ to be a large enough value such that the ADTS can converge. Simulations suggest it is enough to set $\tau^{\text{max}} \geq 5$ when $I \leq 30$ and $K \leq 10$.

  In the data collection and sensing phase of Algorithm \ref{algo:tsg_mu}, each user moves around and completes her claimed sensing tasks. The service provider would only pay a user, if the user has claimed the task in the planning phase, and has completed the task with an acceptable quality. As in Field Agent \cite{fieldagent}, for quality control, the service provider can employ techniques such as GPS marking (for location verification), time stamping (for time verification), and photo/video confirmation. Except the user input (line 2) and sensing execution (lines 16-19), all the other computation, communication, and information checking from maps can be done by the mobile app on behalf of the users.


\begin{algorithm} [t] \small
\caption{\emph{Information Update Algorithm for the Service Provider.}}
\begin{algorithmic} [1] \label{algo:tsg_sp}

\STATE \underline{Initialization}

\STATE Announce the task description, location $l[k]$, time $t[k]$, and reward $\rho[k]$ for all task $k \in \mathcal{K}$ on the mobile app interface.

\STATE Allocate memory for $q_i^{(k,t)}, \, \forall \, i \in \mathcal{I}, (k,t) \in \mathcal{H}$.

\STATE Initialize clock timer $\tau := 1$ on the mobile app interface.

\STATE \underline{Information Update for Task Selection Game in Algorithm \ref{algo:tsg_mu}}

\STATE \textbf{repeat}

\STATE $ \ \ \ $ \textbf{if} task selection update message $\boldsymbol{q}_i$ is received from user $i$

\STATE $ \ \ \ \ \ $ Calculate the number of participants for all $(k,t) \in \mathcal{H}$: \\ \quad\quad\quad $q^{(k,t)} := \sum_{i \in \mathcal{I}} q_i^{(k,t)}, \, \forall \, (k,t) \in \mathcal{H}$.

\STATE $ \ \ \ \ \ $ Update $q^{(k,t)},  \, \forall \, (k,t) \in \mathcal{H}$ on the mobile app interface.

\STATE $ \ \ \ $ \textbf{end if}

\STATE $ \ \ \ $ Set $\tau := \tau + 1$.

\STATE \textbf{until} $\tau \geq \tau^{\text{max}}$.

\end{algorithmic}
\end{algorithm}

\section{Centralized Task Allocation} \label{sec:centralized}

  In this section, we consider the benchmark centralized task allocation (CTA) problem, where the service provider seeks to maximize the social surplus in the TSG.
	We prove that the CTA problem is NP-hard.
  Due to the high complexity of finding the optimal solution, we propose a heuristic greedy centralized algorithm, which turns out to have close-to-optimal performance in our simulations. 

\subsection{Centralized Task Allocation Problem}

  In the CTA benchmark problem, the service provider allocates tasks to the users in order to maximize the social surplus (i.e., users' total rewards minus total movement costs). The social surplus represents the maximum total profit that the service provider can generate, under the ideal case that all the users are under its direct control\opt{opt2}{\footnote{In this case, each user will get a zero payoff, i.e., the payment from the service provider equals to the user's movement costs involved in finishing the tasks.}}. 

  For user $i$'s given strategy (task-time route choice) $r_i$ in \eqref{equ:ri}, we use $y_i^k(r_i) = 1$ to denote that user $i$ works on task $k$ under strategy $r_i$, and $y_i^k(r_i) = 0$ otherwise. That is,
\begin{equation} 
		y_i^k(r_i) =
\begin{cases}
  1,  & \mbox{if $(k,t[k]) \in r_i$},\\
  0, & \mbox{otherwise.}
\end{cases} 
\end{equation}

  Given the users' strategy profile $\boldsymbol{r}$, the social surplus is
\begin{equation} \label{equ:surplus}
	\text{surplus}(\boldsymbol{r}) = \text{reward}(\boldsymbol{r}) - \text{cost}(\boldsymbol{r}),
\end{equation}
where $\text{reward}(\boldsymbol{r})$ is the total rewards received by all users, and $\text{cost}(\boldsymbol{r})$ is the total movement costs of all users. Since the reward $\rho[k]$ is equally shared among all the users working on task $k$, $\text{reward}(\boldsymbol{r})$ equals the total reward of those tasks that have at least one assigned user. So we have
\begin{equation} \label{equ:reward}
	\text{reward}(\boldsymbol{r}) = \sum_{k \in \mathcal{K}} \textbf{1}_{\{\sum_{i \in \mathcal{I}} y_i^{k}(r_i) \geq 1 \}} \rho[k],
\end{equation}
where $\textbf{1}_{\{.\}}$ is the indicator function.
  Moreover, we have
\begin{equation} 
	\text{cost}(\boldsymbol{r}) = \sum_{i \in \mathcal{I}} \text{cost}_i(r_i) = \sum_{i \in \mathcal{I}} \sum_{a=1}^{n-1} c_i ^{l[k_i^a],l[k_i^{a+1}]},
\end{equation}
where $\text{cost}_i(r_i) = \sum_{a=1}^{n-1} c_i ^{l[k_i^a],l[k_i^{a+1}]}$ is the movement cost of user $i$.

  In practice, it is reasonable to assume that the total movement cost of a user is non-decreasing in the number of locations she visited. That is, 
\begin{equation} \label{equ:triangle}
	c_i^{l,l'} + c_i^{l',l''} \geq c_i^{l,l''}, \, \forall \, i \in \mathcal{I}, l, l', l'' \in \mathcal{L}.
\end{equation}

  With this assumption, we can show that there always exists an optimal task allocation where each task is allocated to at most one user.

\begin{lemma} \label{LEM:SOCIALSURPLUS}
  There always exists a social surplus maximizing  (optimal) task assignment $\displaystyle \boldsymbol{r}^* = \!\!\!\! \argmax_{\boldsymbol{r} \in \mathcal{R}_1 \times \ldots \times \mathcal{R}_I} \text{surplus}(\boldsymbol{r})$ such that $\sum_{i \in \mathcal{I}} y_i^{k}(r_i^*) \leq 1, \, \forall \, k \in \mathcal{K}$.
\end{lemma}

The proof of Lemma \ref{LEM:SOCIALSURPLUS} is given in Appendix \ref{app:socialsurplus}.
With this lemma\opt{opt2}{\footnote{Note that there can be other optimal solutions that do not satisfy Lemma \ref{LEM:SOCIALSURPLUS}. However, these solutions can be obtained readily from the optimal solution in Lemma \ref{LEM:SOCIALSURPLUS} as we will discuss in Appendix \ref{sec:multiplesoln}.}}, we can establish the NP-hardness of the CTA benchmark. 

\begin{theorem} \label{THM:NP_HARD}
  The problem of finding the social surplus maximization solution of the TSG is NP-hard.
\end{theorem}	

\opt{opt1}{The proof sketch of Theorem \ref{THM:NP_HARD} is given in Appendix \ref{app:NP_hard}.}
\opt{opt2}{The proof of Theorem \ref{THM:NP_HARD} is given in Appendix \ref{app:NP_hard}.}
Theorem \ref{THM:NP_HARD} motivates us to consider a greedy heuristic algorithm to solve the CTA problem.

\begin{algorithm} [t] \small
\caption{\emph{Greedy Centralized (GC) Algorithm for Task Allocation.}}
\begin{algorithmic} [1] \label{algo:greedy}

\STATE \underline{Initialization}

\STATE Obtain movement costs, speeds, and initial locations from all the users: $c_i^{l,l'}, \, \forall \, l,l' \in \mathcal{L}, i \in \mathcal{I}$, $\nu_i$ and $l_i^{\text{init}}, \, \forall \, i \in \mathcal{I}$.

\STATE Initialize the next available time and current location of the users: $\phi_i := 1$ and $l_i := l_i^{\text{init}}, \, \forall \, i \in \mathcal{I}$.


\STATE Initialize the optimization variables $y_i^k := 0, \, \forall \, i \in \mathcal{I}, k \in \mathcal{K}$.

%
%
%

\STATE \underline{Greedy Centralized Task Allocation}

\STATE Sort the tasks in set $\mathcal{K}$ in the ascending order of the execution time $t[k]$.

\STATE \textbf{for} $k = 1$ to $K$

\STATE $ \ \ \ $ Set $flag := 0$.

\STATE $ \ \ \ $ Sort the users in set $\mathcal{I}$ in the ascending order of $c_i^{l_i, l[k]}$. 

\STATE $ \ \ \ $ Set $i := 1$.

\STATE $ \ \ \ $ \textbf{while} $i \leq I$ \textbf{and} $flag = 0$

\STATE $ \ \ \ \ \ $ \textbf{if} $k \in \mathcal{K}_i$ \textbf{and} $t[k] - \phi_i \geq \Delta_i^{l_i, l[k]}$ \textbf{and} $\rho[k] \geq c_i^{l_i, l[k]}$

\STATE $ \ \ \ \ \ \ \ $ Update user $i$'s next available time $\phi_i := t[k]$ and \\ \quad\quad\quad\quad current location $l_i := l[k]$.

\STATE $ \ \ \ \ \ \ \ $ Indicate the task allocation $y_i^k := 1$ and set $flag := 1$.

\STATE $ \ \ \ \ \ $ \textbf{end if}

\STATE $ \ \ \ \ \ $ Set $i := i + 1$.

\STATE $ \ \ \ $ \textbf{end while}

\STATE \textbf{end for}

\STATE \textbf{for} $i \in \mathcal{I}$

\STATE $ \ \ \ $ Inform user $i$ the task allocation $(y_i^k, \, \forall \, k \in \mathcal{K})$.

\STATE \textbf{end for}

\STATE \underline{Data Collection and Sensing Phase}

\STATE \textbf{for} each user $i \in \mathcal{I}$

\STATE $ \ \ \ $ Move and complete the sensing tasks based on the \\ \quad\quad\quad task allocation $(y_i^k, \, \forall \, k \in \mathcal{K})$.

\STATE \textbf{end for}

\end{algorithmic}
\end{algorithm}

\subsection{Greedy Centralized Task Allocation Algorithm}

  We propose a low complexity greedy centralized (GC) algorithm (Algorithm \ref{algo:greedy}),  which computes an approximate solution to the social surplus maximization problem. The key idea is that the service provider sorts the tasks in ascending order of the execution time, and then allocates one user to each task sequentially in a greedy manner. 
  Specifically, in line 6, the service provider first serves tasks with the earliest execution time.
  In line 9, the service provider sorts users in the order of an increasing movement costs (from the users' current locations to the location of the current task $k$). Line 12 indicates that a user $i$ will be chosen if it is eligible to perform the task ($i:k \in \mathcal{K}_i$), can reach the task location on time ($t[k] - \phi_i \geq \Delta_i^{l_i, l[k]}$), and has the incentive to perform the task ($\rho[k] \geq c_i^{l_i, l[k]}$) (line 12). Here $\phi_i$ is the first time slot of user $i$ after finishing any existing allocating task.
  It is possible that some task $k$ may not be allocated to any users (i.e., $y_i^k = 0, \, \forall \, i \in \mathcal{I}$).

\vspace{0.2cm}
\begin{lemma} \label{LEM:GREEDY}
  Algorithm \ref{algo:greedy} has a computational complexity of $\mathcal{O}(KI \log I)$.
\end{lemma}

The proof of Lemma \ref{LEM:GREEDY} is given in Appendix \ref{app:greedy}.

\section{Performance Evaluations} \label{sec:pe}

\begin{table} [t]
\caption{Simulation Parameters}
\label{table:parameter}
\centering
\begin{tabular}{|c|c|}
\hline
\textbf{Parameters} & \textbf{Values} \\
\hline
Number of tasks $K$ & 10 \\
\hline
Number of time slots $T$ & 15\\
\hline
Rewards for three levels of tasks & $\$10$, $\$15$, $\$20$ \\
\hline
Duration of a time slot $\delta$ & $1$ minute\\
\hline
Movement speed $\nu$ & $0.1$ km/min\\
\hline
\end{tabular}
\end{table}
%

\begin{figure*}[t]
\hspace{-0.5cm}
\centering
\begin{minipage}[t]{0.3\linewidth}
       \includegraphics[width=5.5cm, trim = 0.5cm 0cm 0cm 0cm, clip = true]{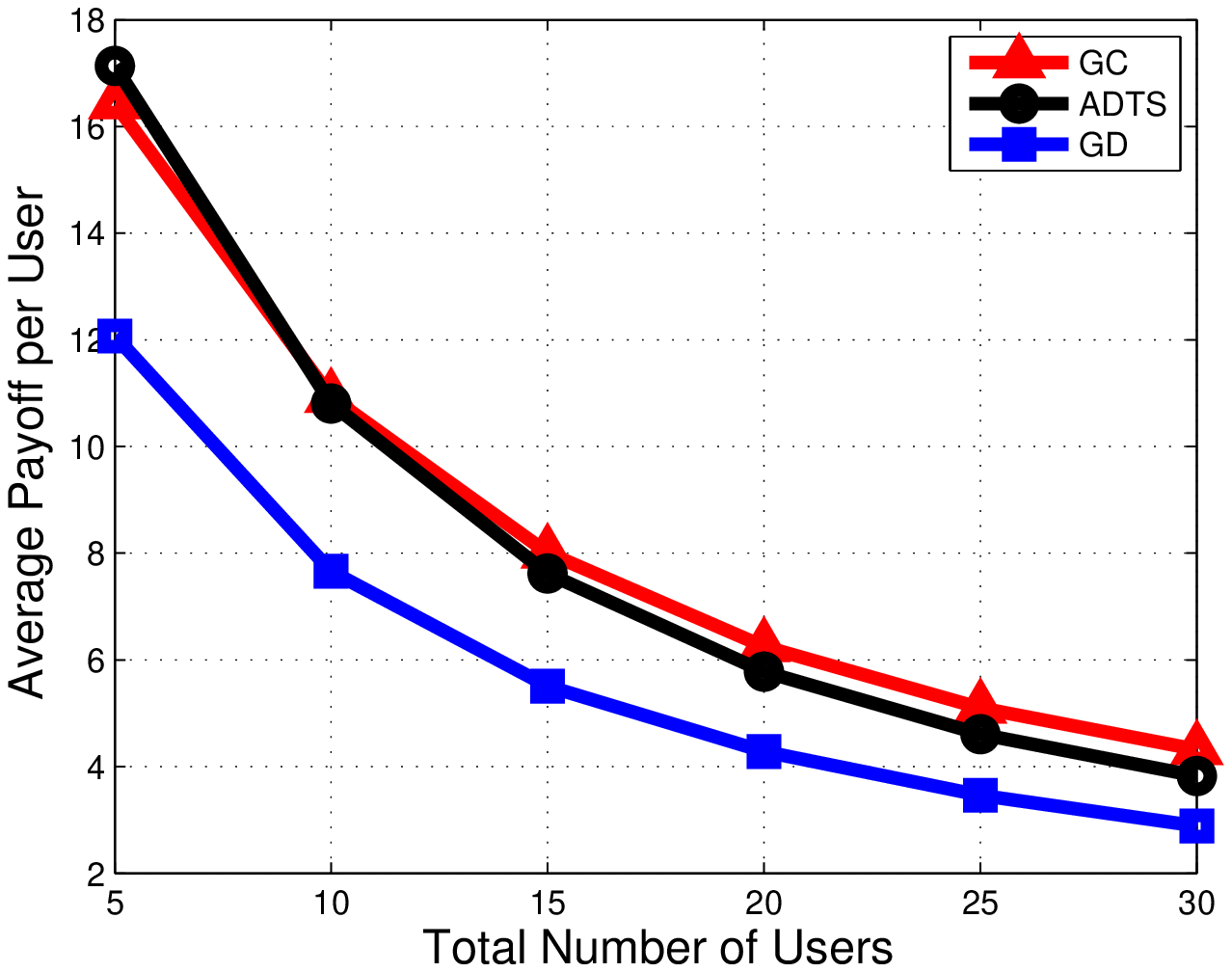}
   \caption{The average user payoff versus the total number of users $I$ for $K = 10$ and $c^{\text{move}} = 0.1$.}
   \label{fig:payoff_numusers}
\end{minipage}
\quad
\begin{minipage}[t]{0.3\linewidth}
       \includegraphics[width=5.5cm, trim = 0.5cm 0cm 0cm 0cm, clip = true]{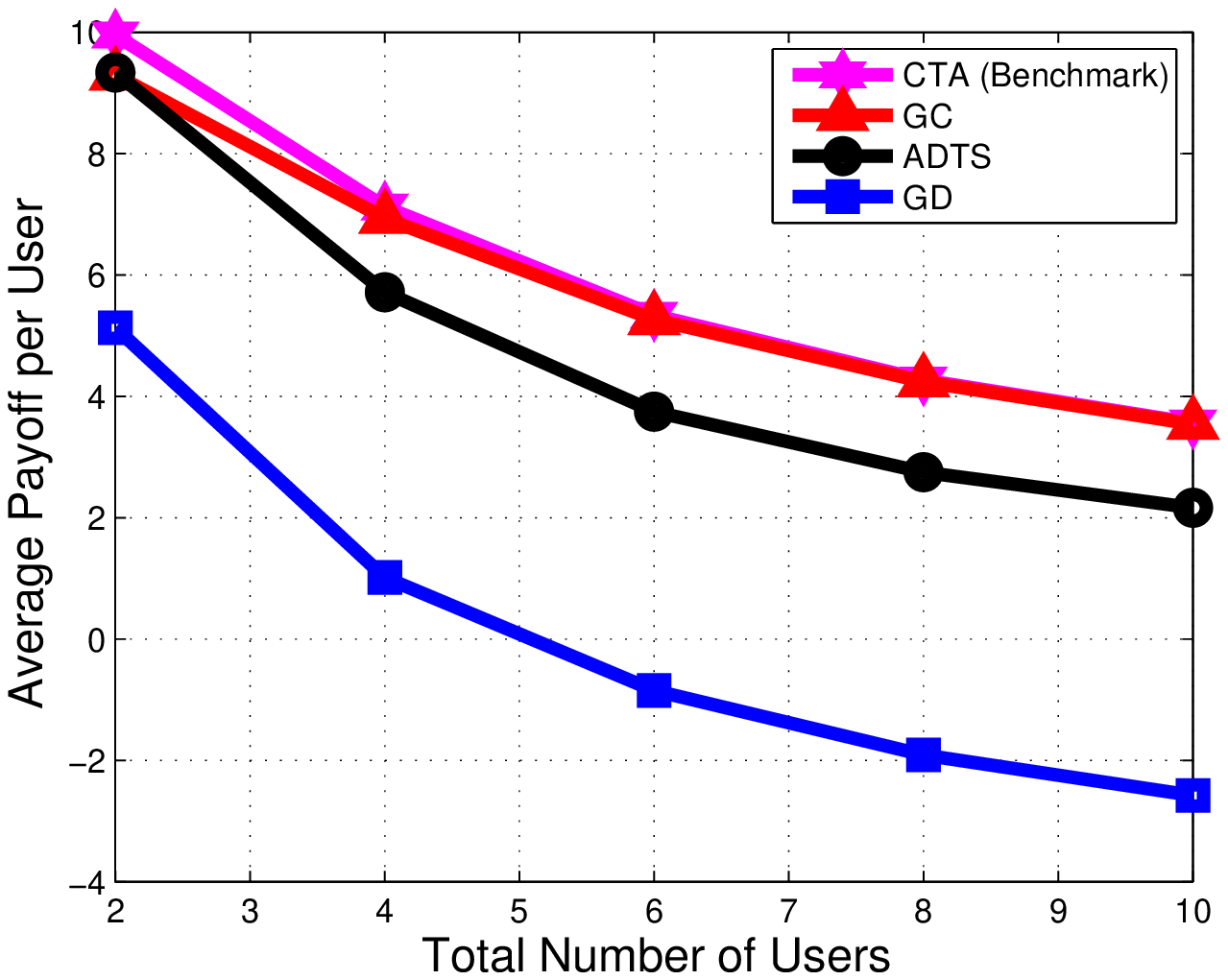}
   \caption{The average user payoff versus the total number of users $I$ for $K = 5$ and $c^{\text{move}} = 1$.}
   \label{fig:payoff_benchnumusers}
\end{minipage}
\quad
\begin{minipage}[t]{0.3\linewidth}
       \includegraphics[width=5.5cm, trim = 0.5cm 0cm 0cm 0cm, clip = true]{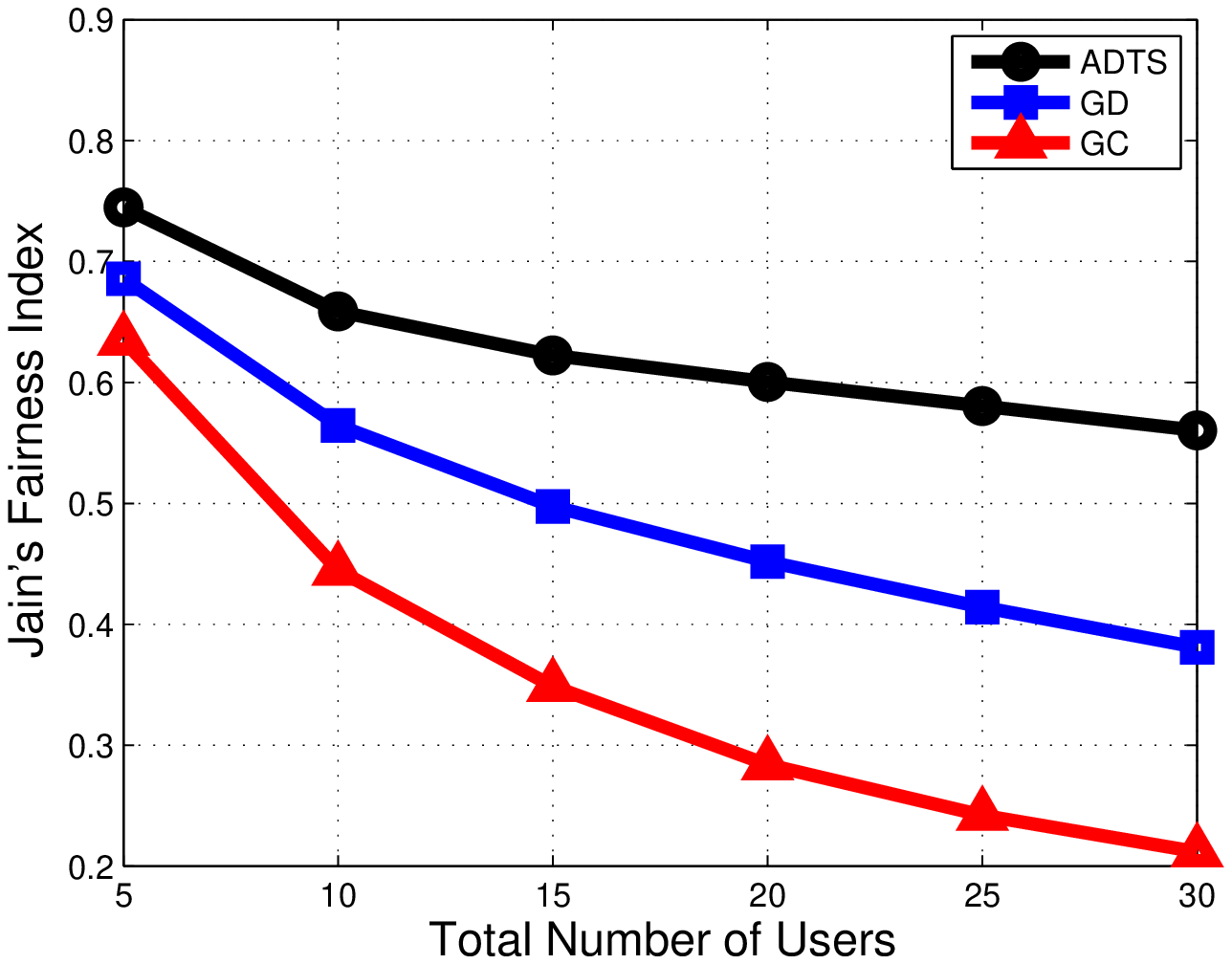}
   \caption{Jain's fairness index versus the total number of users $I$ for $K = 10$ and $c^{\text{move}} = 0.1$.}
   \label{fig:jain_numusers}
\end{minipage}
\end{figure*}

\begin{figure*}[t]
\hspace{-0.5cm}
\centering
\begin{minipage}[t]{0.3\linewidth}
       \includegraphics[width=5.5cm, trim = 0.5cm 0cm 0cm 0cm, clip = true]{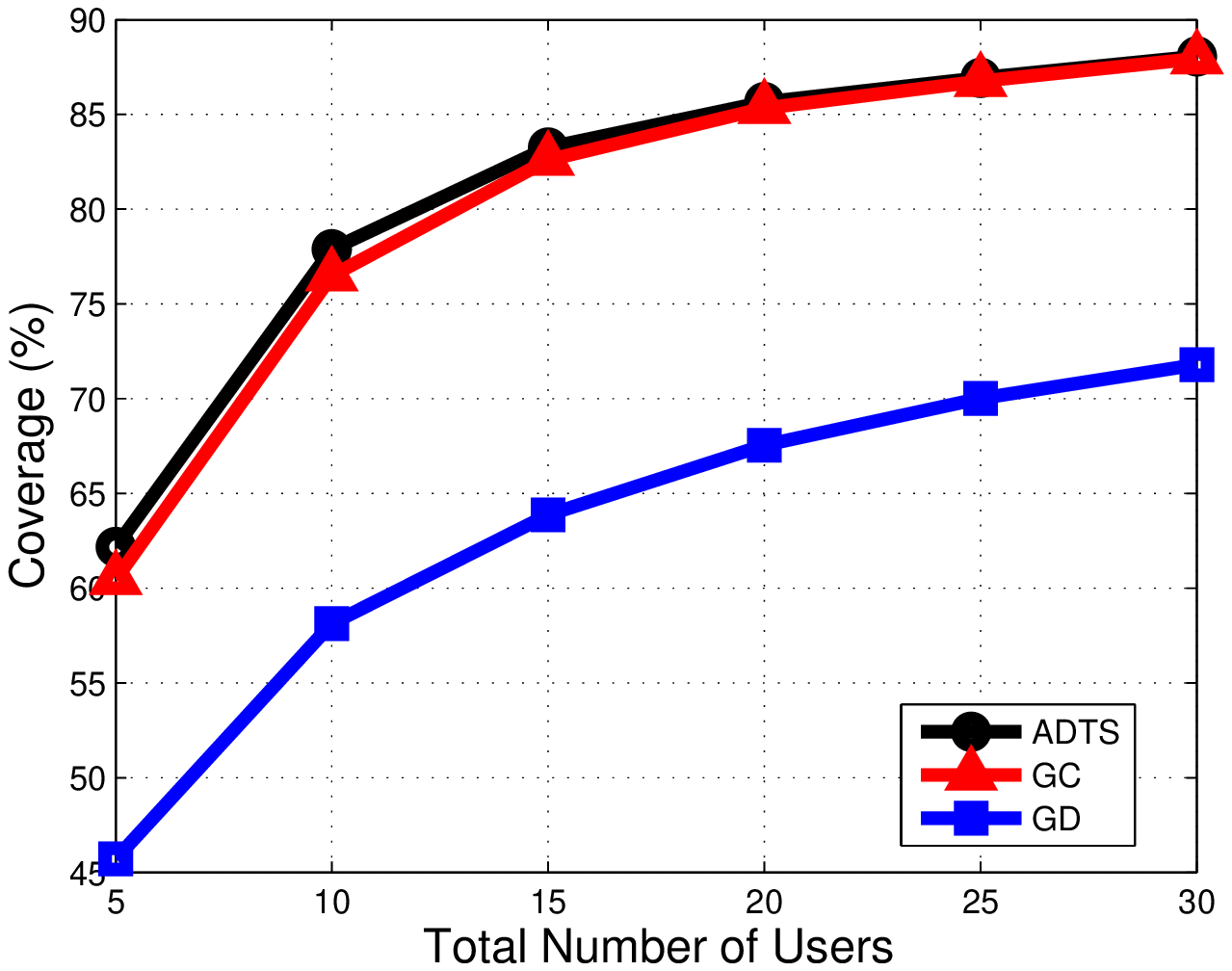}
   \caption{The coverage versus the total number of users $I$ for $K = 10$ and $c^{\text{move}} = 0.1$.}
   \label{fig:coverage_numusers}
\end{minipage}
\quad
\begin{minipage}[t]{0.3\linewidth}
       \includegraphics[width=5.5cm, trim = 0.5cm 0cm 0cm 0cm, clip = true]{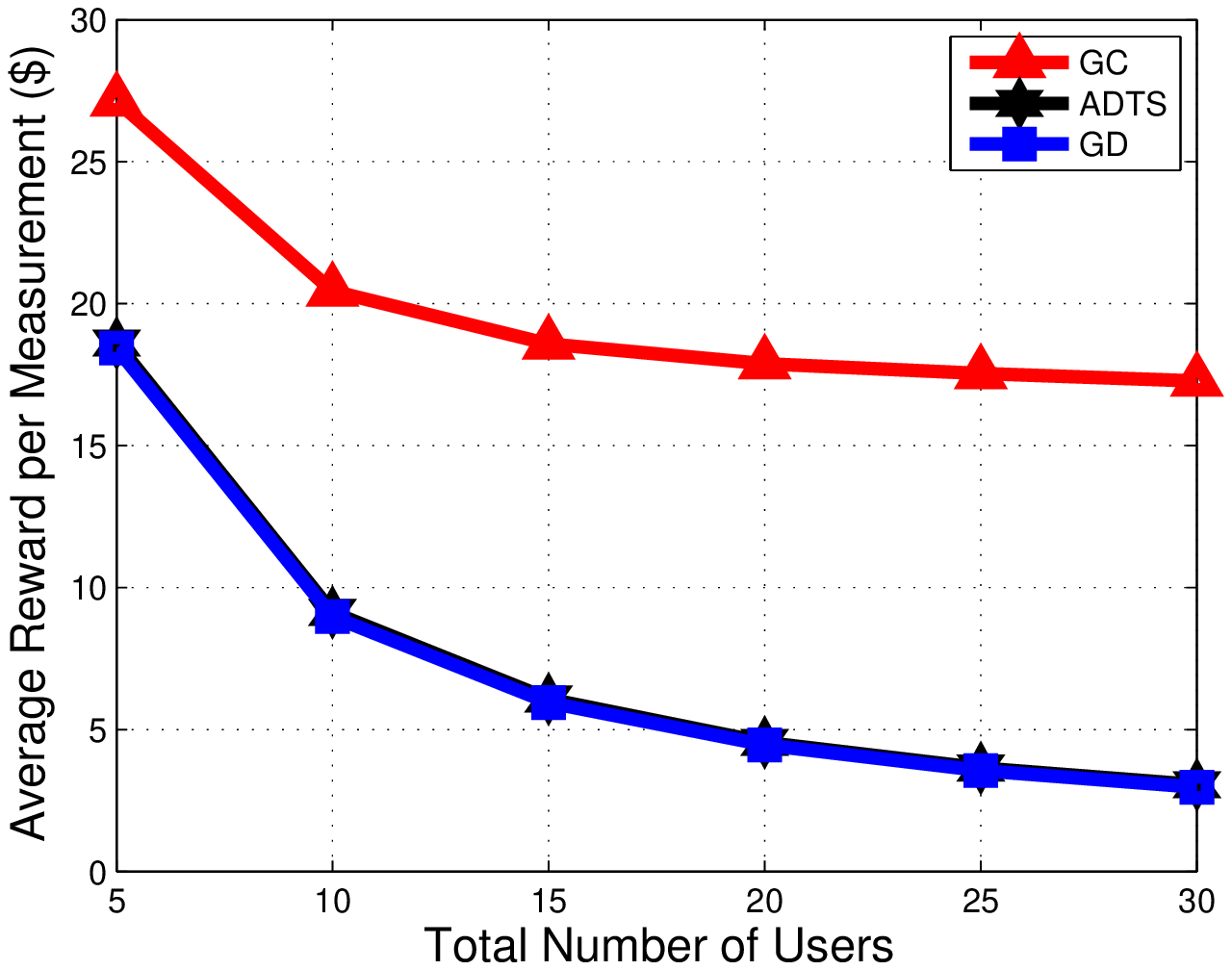}
   \caption{The average reward per measurement versus the total number of users $I$ for $K = 10$ and $c^{\text{move}} = 0.1$.}
   \label{fig:rewardpermeasure_numusers}
\end{minipage}
\quad
\begin{minipage}[t]{0.3\linewidth}
       \includegraphics[width=5.5cm, trim = 0.5cm 0cm 0cm 0cm, clip = true]{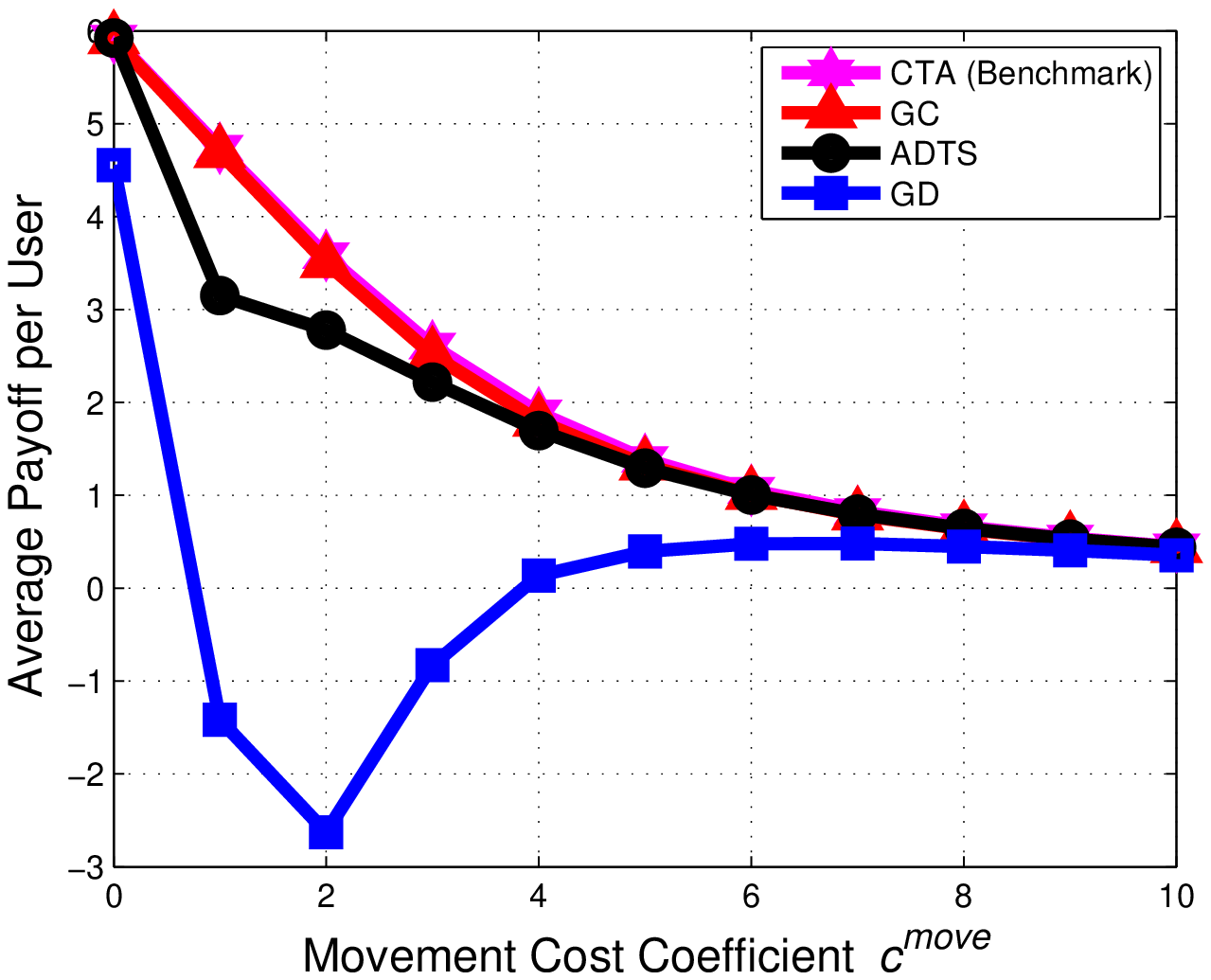} 
   \caption{The average user payoff versus the movement cost coefficient $c^{\text{move}}$ for $I = 7$ and $K = 5$.}
   \label{fig:payoff_benchswcost}
\end{minipage}
\end{figure*}



  In this section, we evaluate the performance of our proposed ADTS scheme by comparing with the CTA benchmark, the GC scheme, and a greedy distributed (GD) scheme. We study the impact of various system parameters on the average user payoff, fairness, coverage, and average reward per measurement on different schemes. We also present an example using real movement time data from Google Maps. Our simulations involve the following schemes:
\begin{itemize}
	\item ADTS scheme: Algorithms \ref{algo:tsg_mu} and \ref{algo:tsg_sp}.	
	\item CTA benchmark: Global optimal solution of the social surplus maximization problem.
	\item GC scheme: Algorithm \ref{algo:greedy} that approximately solves the  social surplus maximization problem.
	\item GD scheme: Each user chooses to work on the earliest feasible task without coordinating with other users. Like the GC scheme, user $i$ should be eligible for the task (i.e., $i:k \in \mathcal{K}_i$), can arrive on time to perform the task (i.e., when $t[k] - \phi_i \geq \Delta_i^{l_i, l[k]}$ given that user $i$ is at location $l_i$),
 and has an incentive to perform sensing assuming no other users share the reward of that task (i.e., $\rho[k] \geq c_i^{l_i, l[k]}$). If there is more than one feasible task with the same execution time, then the user will choose the task with the highest reward.	
\end{itemize}

  For each set of system parameter choices, we run the simulations $1000$ times with randomized initial user locations, and randomized locations and times for the tasks, and show the average value.
  Unless specified otherwise, we assume that the locations of the $K$ tasks and $I$ users are randomly placed in a $1$ km $\times$ $1$ km region.
  Each user is randomly allocated one of three reputation levels, and each task is randomly assigned a threshold. A user is eligible to perform a task when her reputation exceeds the threshold of the task. 
  Each user $i$ moves at constant speed $\nu_i$, and the movement time from location $l \in \mathcal{L}$ to location $l' \in \mathcal{L}$ is $\Delta_i^{{l,l'}} = \left\lceil \frac{\text{dist}(l,l')}{\nu_i \delta} \right\rceil$, where $\text{dist}(l,l')$ is the traveling distance on the road between locations $l$ and $l'$, $\delta$ is the length of a time slot, and $\left\lceil \cdot \right\rceil$ is the ceiling function.
  We assume that the movement cost of user $i$ is linearly proportional to the distance between two locations in the form $c_i^{l,l'} = c_i^{\text{move}} \text{dist}(l,l')$. For simplicity, we assume that all the users have the same movement cost coefficient $c_i^{\text{move}} = c^{\text{move}}$ and the same movement speed $\nu_i = \nu.$ Other simulation parameters are listed in Table \ref{table:parameter}.

\begin{figure*}[t]
\hspace{-0.5cm}
\centering
\begin{minipage}[t]{0.27\linewidth}
       \includegraphics[width=5.5cm, height=4.5cm, trim = 0.5cm 0cm 0cm 0cm, clip = true]{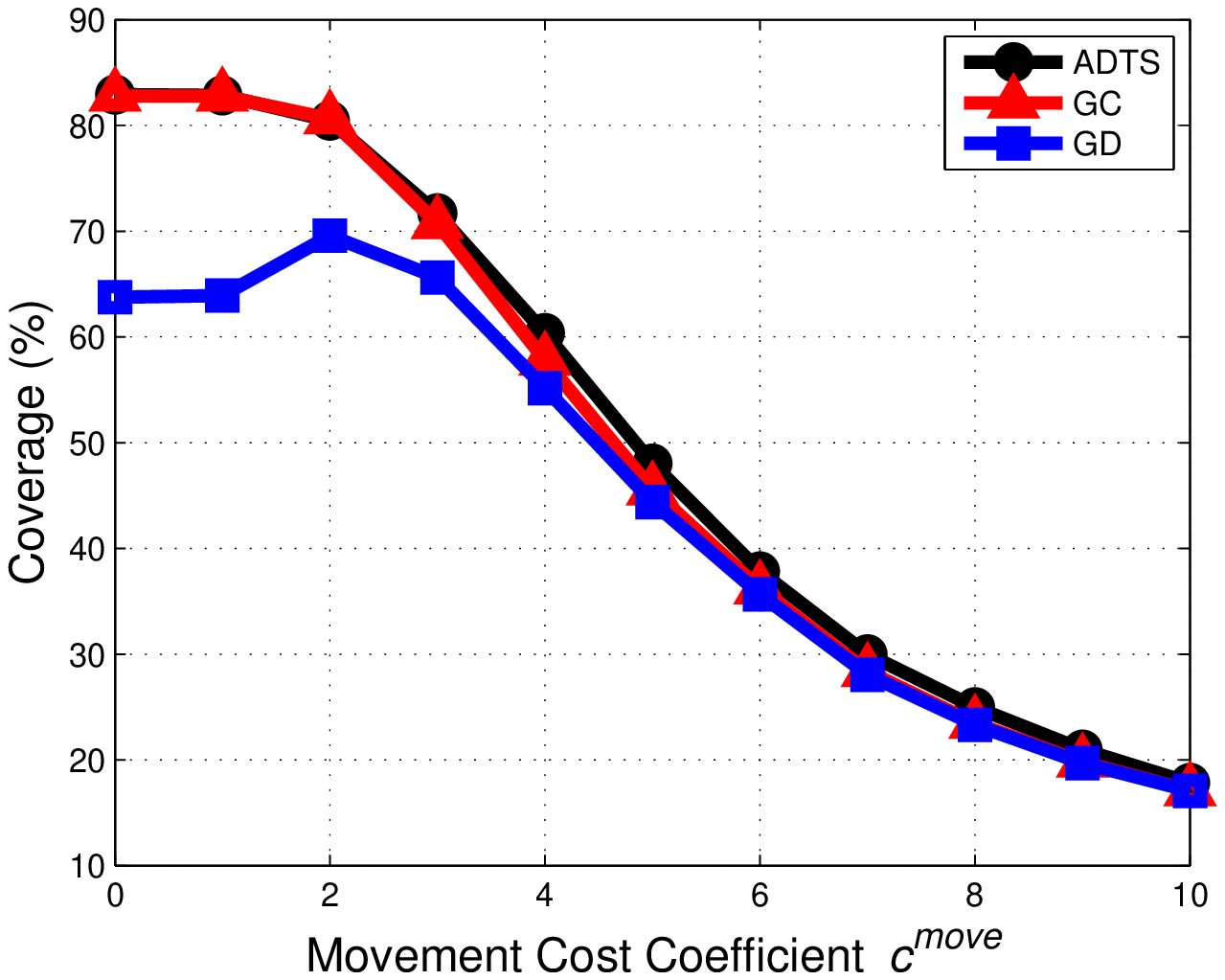}
   \caption{The coverage versus the movement cost coefficient $c^{\text{move}}$ for $I = 10$ and $K = 10$.}
   \label{fig:coverage_swcost}
\end{minipage}
\quad
\begin{minipage}[t]{0.3\linewidth}
       \includegraphics[width=5.5cm, height=4.5cm, trim = 0.5cm 0cm 0cm 0cm, clip = true]{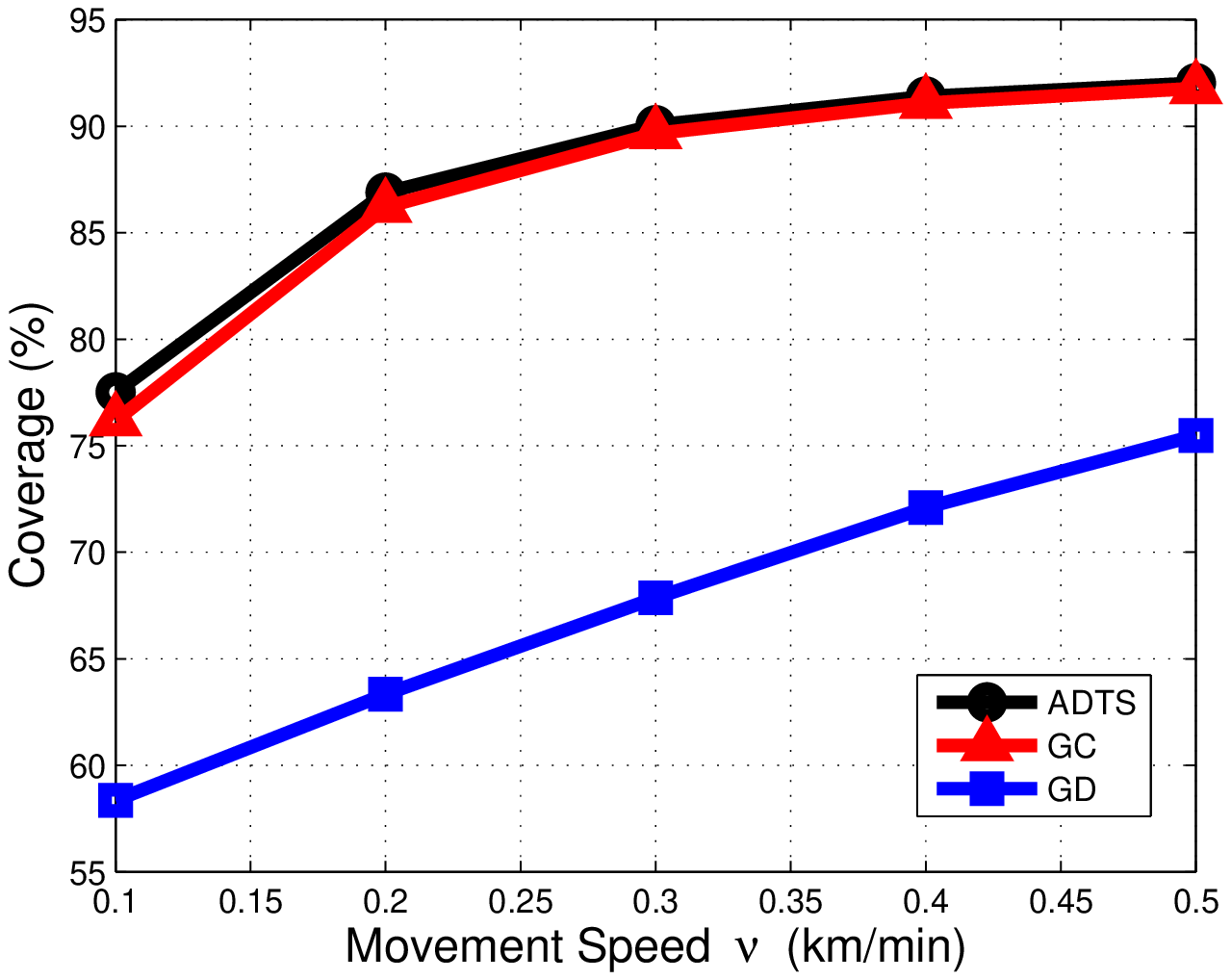}
   \caption{The coverage versus the movement speed $\nu$ for $I = 10$ and $K = 10$.}
   \label{fig:payoff_speed}
\end{minipage}
\quad
\begin{minipage}[t]{0.3\linewidth}
       \includegraphics[width=6cm, trim = 0cm 0cm 0cm 1cm, clip = true]{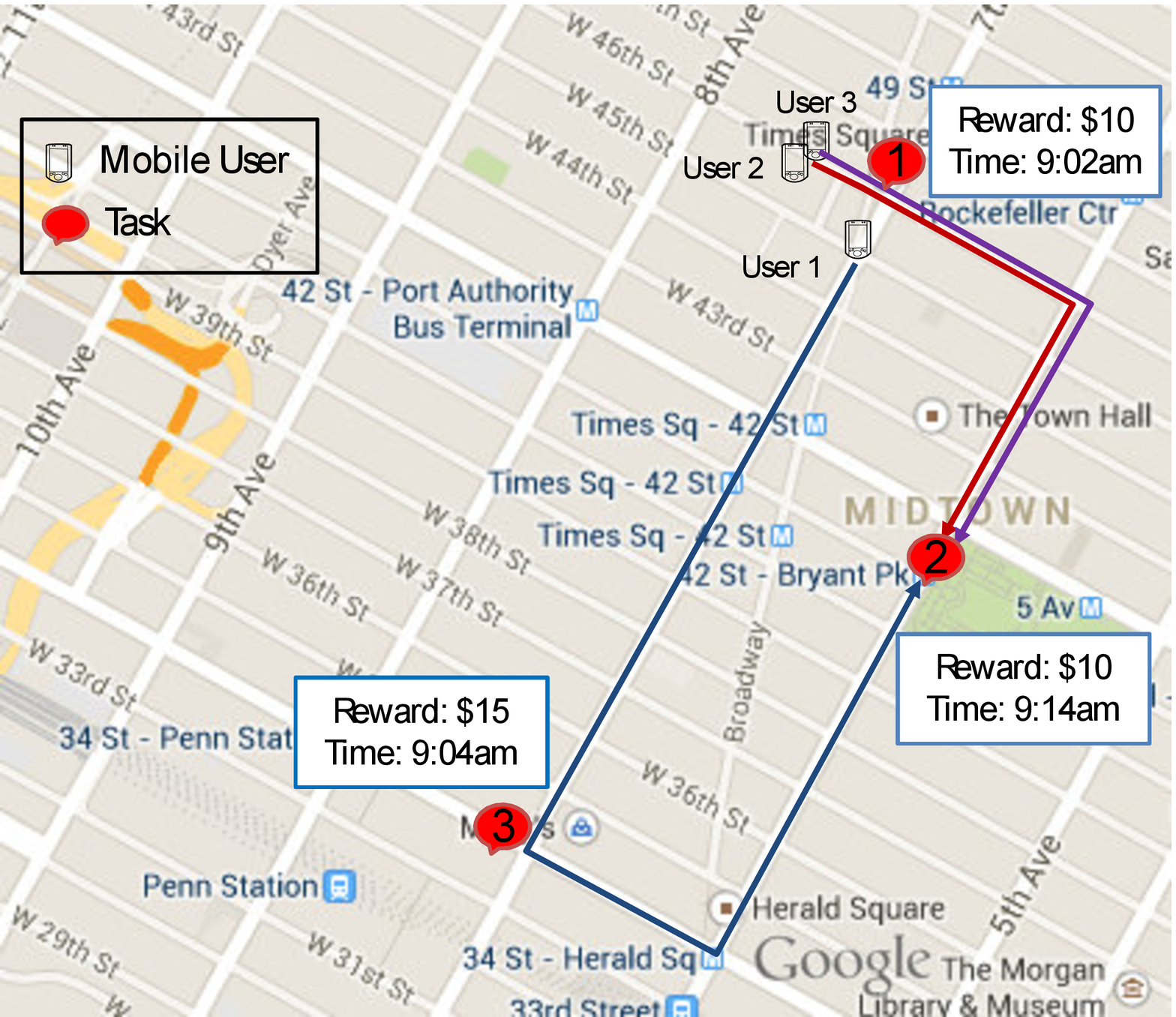}
   \caption{An illustration on the task selection and mobility plans of the users on the map under the ADTS scheme.}
   \label{fig:mapdata}
\end{minipage}
\end{figure*}

\subsection{Average User Payoff and Fairness}

  \textbf{Impact of user number on user payoff:} In Fig.~\ref{fig:payoff_numusers}, we plot the average user payoff against the number of users $I$, with $K = 10$ tasks and movement cost coefficient $c^{\text{move}} = 0.1$. The average user payoff decreases with $I$ under all three schemes, due to the increased competition with the increasing number of users. The ADTS scheme achieves a similar average user payoff as the GC scheme.

  \textbf{Comparison with centralized optimal solution:} In Fig.~\ref{fig:payoff_benchnumusers}, we plot the average user payoff against $I$ in a smaller scale, with $K = 5$ and $c^{\text{move}} = 1$, so that we can evaluate the CTA benchmark as well. Note that GC achieves a similar payoff to the benchmark, especially when $I$ is large.

  \textbf{Fairness:} In Fig.~\ref{fig:jain_numusers}, we study how the payoffs are distributed among the users by plotting Jain's fairness index \cite{jain_aq84} (defined as $\bigl( \sum_{i \in \mathcal{I}} U_i(\boldsymbol{r}) \bigr)^2 / \bigl( I \sum_{i \in \mathcal{I}} U_i(\boldsymbol{r})^2 \bigr)$) against $I$, with $K = 10$ and $c^{\text{move}} = 0.1$. The ADTS scheme has the highest fairness index, as each user has an equal chance to update her strategy profile in ADTS in Algorithm \ref{algo:tsg_mu}. 
   On the other hand, the GC scheme has the lowest fairness index, as it allocates $K$ tasks to at most $K$ users. As a result, when $I$ increases, the number of users not allocated any task increases, so the fairness decreases.
%
\subsection{Coverage and Reward Per Measurement}

  We examine the coverage and average reward per measurement under the three schemes, with $K = 10$ tasks and $c^{\text{move}} = 0.1$.
  The coverage is defined as the percentage of tasks with at least one measurement received.

  \textbf{Impact of user number on coverage:} In Fig.~\ref{fig:coverage_numusers}, we plot the coverage against $I$. When $I$ increases, we see the coverage under all the schemes increases, as there are more users to work on the tasks. The ADTS scheme has a slightly higher coverage than the GC scheme, as users try to avoid too much overlapping with each other in ADTS. The large difference between the coverage of the ADTS and GD schemes is due to the lack of coordination in the GD scheme.

  \textbf{Average reward per measurement:} Fig.~\ref{fig:rewardpermeasure_numusers} shows the average reward per measurement against $I$. We can see that the ADTS and GD schemes have the same smaller average reward per measurement than the centralized scheme GC. From a service provider's point of view, with the same amount of total reward, lower average reward per measurement means that more sensing data can be collected, which is beneficial for the service provider.


\subsection{Impact of Moving Cost and Speed}

\begin{table} [t]
\caption{Movement Time Data from Google Maps} %
\label{table:swtime}
\centering
\begin{tabular}{|c|c|c|c|}
\hline
  \small{\textbf{Drive/Walk (min)}} & \small{Location $1$} & \small{Location $2$} & \small{Location $3$}\\
\hline
  \small{Location $1$} & 0 / 0 & 3 / 11 & 4 / 15 \\
\hline
  \small{Location $2$} & 3 / 11 & 0 / 0 & 2 / 6 \\
\hline
  \small{Location $3$} & 4 / 15 & 2 / 6 & 0 / 0 \\
\hline
\end{tabular}
\end{table}




  \textbf{Impact of movement cost on user payoff:} In Fig.~\ref{fig:payoff_benchswcost}, we plot the average user payoff against $c^{\text{move}}$ with $I = 7$ and $K = 5$.
   As $c^{\text{move}}$ increases, the payoffs for collecting measurements under both ADTS and GC decrease.  Moreover, since users under both ADTS and GC are aware of the actual reward that they can receive (after sharing with other users), the users would not receive a negative payoff for collecting data.
  However, for the GD scheme without any user coordination, a user makes decisions with the belief that she will gain the full reward $\rho[k]$ for task $k$, rather than the actual reward she can achieve (after sharing with other users). When $c^{\text{move}} \leq 4$, users become overly aggressive in GD, and it becomes common that multiple users work on the same task. As a result, the actual reward per user is much less than the full reward of the task, which results in the decrease of the average payoff (even to a negative value) in GD.
   However, when $c^{\text{move}}$ increases beyond $4$, users are less likely to work on the same task in GD, so the evaluation of the reward is more accurate, and the average user payoff in GD gradually converges to a non-negative value, similar to those under the ADTS and GC schemes.

  \textbf{Impact of movement cost on coverage:} In Fig.~\ref{fig:coverage_swcost}, we plot the coverage against the movement cost coefficient $c^{\text{move}}$ with $I = 10$ and $K = 10$. As $c^{\text{move}}$ increases, the coverage decreases in general, because the users are less willing to perform sensing. The slight increase in coverage in GD for $c^{\text{move}} \leq 2$ is due to the inaccurate reward estimation mentioned in the previous paragraph.


  \textbf{Impact of movement speed on coverage:} In Fig.~\ref{fig:payoff_speed}, we plot the coverage against the movement speed $\nu$ for $I = 10$ and $K = 10$. As $\nu$ increases, the coverages under all three schemes increases. The ADTS scheme has a slightly larger coverage than the GC scheme.

\subsection{A Real-World Example Using Google Maps}

\begin{table} [t]
\caption{Summary of Performance in the Real-world Example} 
\label{table:realworld}
\centering
\begin{tabular}{|c|c|c|c|}
\hline
   & \small{\textbf{ADTS}} & \small{\textbf{GC} and \textbf{CTA}}  & \small{\textbf{GD}} \\
   & & \small{\textbf{benchmark}} & \\
\hline
 \small{\textbf{Average user payoff}} & \$10.33 & \$11 & \$7 \\
\hline
 \small{\textbf{Jain's fairness index}} & 0.93 & 0.64 & 0.93 \\
\hline
  \small{\textbf{Coverage}} & 100\% & 100\% & 66.67\% \\
\hline
\end{tabular}
\end{table}

  To better illustrate the schemes, we next introduce a more concrete example based on the neighborhood map in New York City with real data from Google Maps. We consider $I = 3$ users and $K = 3$ tasks.
    The locations of the users, tasks, the execution times, and rewards are shown in Fig.~\ref{fig:mapdata}.  
    We assume that user $1$ prefers driving, while users $2$ and $3$ prefer walking. We compute the movement time of driving and walking between different locations through Google Maps, as shown in Table \ref{table:swtime}.
	For the movement cost, we assume that driving between any two different locations costs $\$2$ and walking is free. 
	
Fig.~\ref{fig:mapdata} shows the task selection plans of three users (i.e., the blue, red, and purple broken lines with arrows)  under the ADTS scheme. As we can see, user $1$ takes the task selection plan of task $3 \rightarrow$ task $2$, while users $2$ and $3$ take the task selection plan of task $1 \rightarrow$ task $2$.
  The intuition is that as users $2$ and $3$ are pedestrians, who do not have enough time to go to the far away location $3$ from their initial locations. As a result, user $1$, who is driving, prefers to take task $3$ with the minimum competition, and enjoys a higher reward.
  After finishing their first chosen task, all three users have enough time to work on task $2$, and share the reward of $\$10$.
  In contrast, under both the GC scheme and CTA benchmark (not shown in the figure), user $1$ works on task $3$, user $2$ works on both tasks $1$ and $2$, and user $3$ does not have any task to work on.
  For the GD scheme (not shown in the figure), since the initial locations of all the users are nearby, their task selection plans are identical: task $1 \rightarrow$ task $2$, as everyone chooses to pursue the earliest task without any coordination.
  The performance of the three schemes are summarized in Table \ref{table:realworld}. We can see that the ADTS scheme achieves a high level of fairness, coverage, and user payoff.

\section{Conclusion} \label{sec:concl}

  Motivated by commercial mobile crowdsensing applications, such as Gigwalk and Field Agent, we studied the distributed time-sensitive and location-dependent task selection problem in this paper.
  We examined the problem from the perspectives of both noncooperative game and centralized optimization.  
	We proposed an asynchronous and distributed task selection (ADTS) algorithm to help users determine their task selection and mobility plans in a distributed fashion, based on limited information on users' aggregate task choices. We proved that finding the social surplus maximization solution is NP-hard. 
  Simulation results showed that the ADTS scheme achieves the highest Jain's fairness index and coverage as compared with two heuristic schemes.
	In this paper, we focus on the distributed task selections of a fixed set of users. In the future work, we will consider the case a changing user population, where users can dynamically join and leave the system.

\rev{
\section{Acknowledgment}

  This work is supported by University of Macau Grant MYRG2014-00140-FST.
	This work is also supported by the General Research Funds (Project Number CUHK 412713 and 14202814) established under the University Grant Committee of the Hong Kong Special Administrative Region, China.
}

\appendix

%
\opt{opt1}{ 
\section{Proof Sketch of Theorem \ref{THM:FIP}} \label{app:fip}

  The key idea is to show that the mapping
 $$\Phi\left(\boldsymbol{r}\right)=\left(\sum_{\left(k,t\right)\in\mathcal{K}\times\mathcal{T}}\sum_{q=1}^{m^{\left(k,t\right)}\left(\boldsymbol{r}\right)}\frac{\rho^{*}\left[k,t\right]}{q}\right)-\sum_{i\in\mathcal{I}}\sum_{e\in\mathbb{E}\left(r_{i}\right)}g_{i}(e)$$
is a \emph{potential function} \cite{monderer_pg96} of the TSG (where $\mathbb{E}(r_i)$ is defined in Definition \ref{z:ttp points}). This means that
for each strategy profile $\boldsymbol{r}=\left(r_{1},\ldots,r_{I}\right)\in\mathcal{R}_{1}\times\ldots\times\mathcal{R}_{I},$ each user $i\in\mathcal{I},$ and each strategy $r_{i}'\in\mathcal{R}_{i}$
available to user $i$, we have $\Phi\left(r_{i}',\boldsymbol{r}_{-i}\right)-\Phi\left(\boldsymbol{r}\right)=U_{i}\left(r_{i}',\boldsymbol{r}_{-i}\right)-U_{i}\left(\boldsymbol{r}\right).$
 In \cite{monderer_pg96}, the authors prove that every finite game with a potential function has the FIP. Since the TSG is a finite game, we prove the statement in Theorem \ref{THM:FIP}. Full details of the proof can be found in \cite{cheung_dt15_arxiv}. \hfill \QEDclosed 

}
\opt{opt2}{
\section{Proof of Theorem \ref{THM:FIP}} \label{app:fip}

  The key idea is to show that the mapping
 $$\Phi\left(\boldsymbol{r}\right)=\left(\sum_{\left(k,t\right)\in\mathcal{K}\times\mathcal{T}}\sum_{q=1}^{m^{\left(k,t\right)}\left(\boldsymbol{r}\right)}\frac{\rho^{*}\left[k,t\right]}{q}\right)-\sum_{i\in\mathcal{I}}\sum_{e\in\mathbb{E}\left(r_{i}\right)}g_{i}(e)$$
is a \emph{potential function} \cite{monderer_pg96} of the TSG (where $\mathbb{E}(r_i)$ is defined in Definition \ref{z:ttp points}). This means that
for each strategy profile $\boldsymbol{r}=\left(r_{1},\ldots,r_{I}\right)\in\mathcal{R}_{1}\times\ldots\times\mathcal{R}_{I},$ each user $i\in\mathcal{I},$ and each strategy $r_{i}'\in\mathcal{R}_{i}$
available to user $i$, we have $\Phi\left(r_{i}',\boldsymbol{r}_{-i}\right)-\Phi\left(\boldsymbol{r}\right)=U_{i}\left(r_{i}',\boldsymbol{r}_{-i}\right)-U_{i}\left(\boldsymbol{r}\right).$
 In \cite{monderer_pg96}, the authors prove that every finite game with a potential function has the FIP. Since the TSG is a finite game, we prove the statement in Theorem \ref{THM:FIP}.

 To show that $\Phi\left(\boldsymbol{r}\right)$ is a potential function, we shall separate our sums into two parts.
\begin{equation}
\Phi^{V}\left(\boldsymbol{r}\right)=\sum_{\left(k,t\right)\in\mathcal{K}\times\mathcal{T}}\sum_{q=1}^{m^{\left(k,t\right)}\left(\boldsymbol{r}\right)}\frac{\rho^{*}\left[k,t\right]}{q}\label{eq:potV}
\end{equation}
and
\begin{equation}
\Phi^{E}\left(\boldsymbol{r}\right)=\sum_{i\in\mathcal{I}}\sum_{e\in\mathbb{E}\left(r_{i}\right)}-g_{i}(e).\label{eq:potE}
\end{equation}
Adding equations \eqref{eq:potV} and \eqref{eq:potE} gives
\begin{equation}
\Phi\left(\boldsymbol{r}\right)=\Phi^{V}\left(\boldsymbol{r}\right)+\Phi^{E}\left(\boldsymbol{r}\right).\label{eq:potadd}
\end{equation}

Also, for a subset $S\subseteq\mathcal{K}\times\mathcal{T}$ of task-time
points, let us define
\begin{equation}
\Phi_{S}^{V}\left(\boldsymbol{r}\right)=\sum_{\left(k,t\right)\in S}\sum_{q=1}^{m^{\left(k,t\right)}\left(\boldsymbol{r}\right)}\frac{\rho^{*}\left[k,t\right]}{q}. \label{eq:potVS}
\end{equation}

Clearly $\Phi_{\mathcal{K}\times\mathcal{T}}^{V}\left(\boldsymbol{r}\right)=\Phi^{V}\left(\boldsymbol{r}\right).$
Let us define
\begin{equation}
U_{i}^{V}\left(\boldsymbol{r}\right)=\sum_{\left(k,t\right)\in\mathbb{V}\left(r_{i}\right)}\frac{\rho^{*}\left[k,t\right]}{m^{\left(k,t\right)}\left(\boldsymbol{r}\right)}\label{eq:UV}
\end{equation}
and
\begin{equation}
U_{i}^{E}\left(\boldsymbol{r}\right)=\sum_{e\in\mathbb{E}\left(r_{i}\right)}-g_{i}(e).\label{eq:UE}
\end{equation}

By comparing this with equation \eqref{eq:potE} one can see that
\begin{equation}
\Phi^{E}\left(\boldsymbol{r}\right)=\sum_{i\in\mathcal{I}}U_{i}^{E}\left(\boldsymbol{r}\right).\label{eq:potE-1}
\end{equation}

Adding equations \eqref{eq:UV} and \eqref{eq:UE} gives
\begin{equation}
U_{i}\left(\boldsymbol{r}\right)=U_{i}^{V}\left(\boldsymbol{r}\right)+U_{i}^{E}\left(\boldsymbol{r}\right). \label{eq:Uadd}
\end{equation}

Suppose that our game is in strategy profile $\boldsymbol{r}=\left(r_{1},..,r_{I}\right)$
and then user $j\in\mathcal{I}$ changes their strategy to $r_{j}'\in\mathcal{R}_{j}$,
and so we have a new strategy profile $\boldsymbol{r}'=\left(r_{j}',\boldsymbol{r}_{-j}\right).$
In order to show that $\Phi$ is an exact potential function we will
show
\begin{equation}
\Phi\left(r_{j}',\boldsymbol{r}_{-j}\right)-\Phi\left(\boldsymbol{r}\right)=U_{j}\left(r_{j}',\boldsymbol{r}_{-j}\right)-U_{j}\left(\boldsymbol{r}\right). \label{eq:pot goal}
\end{equation}

To begin, note that we can expand out the left hand side of equation
\eqref{eq:pot goal} using equation \eqref{eq:potadd} to obtain:
\begin{equation}
\begin{split}
  \Phi\left(r_{j}',\boldsymbol{r}_{-j}\right)-\Phi\left(\boldsymbol{r}\right) \quad\quad\quad\quad\quad\quad\quad\quad\quad\quad\quad\quad\quad\quad \\ 
  =\left[\Phi^{V}\left(r_{j}',\boldsymbol{r}_{-j}\right)-\Phi^{V}\left(\boldsymbol{r}\right)\right]+\left[\Phi^{E}\left(r_{j}',\boldsymbol{r}_{-j}\right)-\Phi^{E}\left(\boldsymbol{r}\right)\right].
\end{split}
\label{eq:lhs expand}
\end{equation}
%

Equation \eqref{eq:potE-1} implies that
\begin{equation}
\begin{split}
\Phi^{E}\left(r_{j}',\boldsymbol{r}_{-j}\right)-\Phi^{E}\left(\boldsymbol{r}\right)=\sum_{i\in\mathcal{I}}\left[U_{i}^{E}\left(r_{j}',\boldsymbol{r}_{-j}\right)-U_{i}^{E}\left(\boldsymbol{r}\right)\right] \\
  =U_{j}^{E}\left(r_{j}',\boldsymbol{r}_{-j}\right)-U_{j}^{E}\left(\boldsymbol{r}\right). \hspace{0.7cm}
\end{split}
\label{eq:eqE}
\end{equation}
%

Using equation \eqref{eq:potVS}, we can obtain that
\begin{equation}
\begin{split}
   \Phi^{V}\left(r_{j}',\boldsymbol{r}_{-j}\right)-\Phi^{V}\left(\boldsymbol{r}\right) \hspace{5cm} \\
 = \left[\Phi_{\mathcal{K}\times\mathcal{T}-\left(\mathbb{V}\left(r_{j}\right)\cup\mathbb{V}\left(r_{j}'\right)\right)}^{V}\left(r_{j}',\boldsymbol{r}_{-j}\right)-\Phi_{\mathcal{K}\times\mathcal{T}-\left(\mathbb{V}\left(r_{j}\right)\cup\mathbb{V}\left(r_{j}'\right)\right)}^{V}\left(\boldsymbol{r}\right)\right] \\
  +\left[\Phi_{\mathbb{V}\left(r_{j}\right)\cap\mathbb{V}\left(r_{j}'\right)}^{V}\left(r_{j}',\boldsymbol{r}_{-j}\right)-\Phi_{\mathbb{V}\left(r_{j}\right)\cap\mathbb{V}\left(r_{j}'\right)}^{V}\left(\boldsymbol{r}\right)\right] \hspace{1.5cm} \\
  +\left[\Phi_{\mathbb{V}\left(r_{j}'\right)-\mathbb{V}\left(r_{j}\right)}^{V}\left(r_{j}',\boldsymbol{r}_{-j}\right)-\Phi_{\mathbb{V}\left(r_{j}'\right)-\mathbb{V}\left(r_{j}\right)}^{V}\left(\boldsymbol{r}\right)\right] \hspace{1.5cm} \\
  +\left[\Phi_{\mathbb{V}\left(r_{j}\right)-\mathbb{V}\left(r_{j}'\right)}^{V}\left(r_{j}',\boldsymbol{r}_{-j}\right)-\Phi_{\mathbb{V}\left(r_{j}\right)-\mathbb{V}\left(r_{j}'\right)}^{V}\left(\boldsymbol{r}\right)\right]. \hspace{1.4cm}
\end{split} \label{eq:bigV}
\end{equation}
%

Now since $m^{\left(k,t\right)}\left(\boldsymbol{r}\right)=m^{\left(k,t\right)}\left(r_{j}',\boldsymbol{r}_{-j}\right)$
holds true for every $(k,t)\in\mathcal{K}\times\mathcal{T}-\left(\mathbb{V}\left(r_{j}\right)\cup\mathbb{V}\left(r_{j}'\right)\right)$
or $(k,t)\in\mathbb{V}\left(r_{j}\right)\cap\mathbb{V}\left(r_{j}'\right)$
we have that 
\begin{equation*}
\begin{split}
  \left[\Phi_{\mathcal{K}\times\mathcal{T}-\left(\mathbb{V}\left(r_{j}\right)\cup\mathbb{V}\left(r_{j}'\right)\right)}^{V}\left(r_{j}',\boldsymbol{r}_{-j}\right)-\Phi_{\mathcal{K}\times\mathcal{T}-\left(\mathbb{V}\left(r_{j}\right)\cup\mathbb{V}\left(r_{j}'\right)\right)}^{V}\left(\boldsymbol{r}\right)\right] \\
	+\left[\Phi_{\mathbb{V}\left(r_{j}\right)\cap\mathbb{V}\left(r_{j}'\right)}^{V}\left(r_{j}',\boldsymbol{r}_{-j}\right)-\Phi_{\mathbb{V}\left(r_{j}\right)\cap\mathbb{V}\left(r_{j}'\right)}^{V}\left(\boldsymbol{r}\right)\right]=0,
\end{split}
\end{equation*}
%
and so equation \eqref{eq:bigV} can be simplified to give
\begin{equation}
\begin{split}
  \Phi^{V}\left(r_{j}',\boldsymbol{r}_{-j}\right)-\Phi^{V}\left(\boldsymbol{r}\right) \hspace{3cm} \\
	= \left[\Phi_{\mathbb{V}\left(r_{j}'\right)-\mathbb{V}\left(r_{j}\right)}^{V}\left(r_{j}',\boldsymbol{r}_{-j}\right)-\Phi_{\mathbb{V}\left(r_{j}'\right)-\mathbb{V}\left(r_{j}\right)}^{V}\left(\boldsymbol{r}\right)\right] \quad \\ 
  + \left[\Phi_{\mathbb{V}\left(r_{j}\right)-\mathbb{V}\left(r_{j}'\right)}^{V}\left(r_{j}',\boldsymbol{r}_{-j}\right)-\Phi_{\mathbb{V}\left(r_{j}\right)-\mathbb{V}\left(r_{j}'\right)}^{V}\left(\boldsymbol{r}\right)\right].\label{eq:simplerV}
\end{split}
\end{equation}
%

Using equation \eqref{eq:potVS}, together with the fact that $m^{\left(k,t\right)}\left(r_{j}',\boldsymbol{r}_{-j}\right)=m^{\left(k,t\right)}\left(\boldsymbol{r}\right)+1,$
$\forall(k,t)\in\mathbb{V}\left(r_{j}'\right)-\mathbb{V}\left(r_{j}\right)$
gives us that
\begin{equation}
\begin{split}
  \Phi_{\mathbb{V}\left(r_{j}'\right)-\mathbb{V}\left(r_{j}\right)}^{V}\left(r_{j}',\boldsymbol{r}_{-j}\right)-\Phi_{\mathbb{V}\left(r_{j}'\right)-\mathbb{V}\left(r_{j}\right)}^{V}\left(\boldsymbol{r}\right) \hspace{2.6cm} \\
 = \hspace{-0.3cm} \sum_{\left(k,t\right)\in\mathbb{V}\left(r_{j}'\right)-\mathbb{V}\left(r_{j}\right)} \!\! \left[\left[\sum_{q=1}^{m^{\left(k,t\right)}\left(r_{j}',\boldsymbol{r}_{-j}\right)} \hspace{-0.2cm} \frac{\rho^{*}\left[k,t\right]}{q}\right] \!\!- \!\! \left[\sum_{q=1}^{m^{\left(k,t\right)}\left(\boldsymbol{r}\right)} \hspace{-0.2cm} \frac{\rho^{*}\left[k,t\right]}{q}\right]\right] \\
  =\sum_{\left(k,t\right)\in\mathbb{V}\left(r_{j}'\right)-\mathbb{V}\left(r_{j}\right)}\frac{\rho^{*}\left[k,t\right]}{m^{\left(k,t\right)}\left(\boldsymbol{r}\right)+1}. \hspace{3.9cm} 
\end{split} \label{eq:smpleradd}
\end{equation}
%

Similarly, using equation \eqref{eq:potVS}, together with the fact
that $m^{\left(k,t\right)}\left(r_{j}',\boldsymbol{r}_{-j}\right)=m^{\left(k,t\right)}\left(\boldsymbol{r}\right)-1,$
$\forall(k,t)\in\mathbb{V}\left(r_{j}\right)-\mathbb{V}\left(r_{j}'\right)$
gives us that
\begin{equation}
\begin{split}
  \Phi_{\mathbb{V}\left(r_{j}'\right)-\mathbb{V}\left(r_{j}\right)}^{V}\left(r_{j}',\boldsymbol{r}_{-j}\right)-\Phi_{\mathbb{V}\left(r_{j}'\right)-\mathbb{V}\left(r_{j}\right)}^{V}\left(\boldsymbol{r}\right) \hspace{3cm} \\
  = \hspace{-0.3cm} \sum_{\left(k,t\right)\in\mathbb{V}\left(r_{j}\right)-\mathbb{V}\left(r_{j}'\right)}\left[\left[\sum_{q=1}^{m^{\left(k,t\right)}\left(r_{j}',\boldsymbol{r}_{-j}\right)} \hspace{-0.2cm} \frac{\rho^{*}\left[k,t\right]}{q}\right]-\left[\sum_{q=1}^{m^{\left(k,t\right)}\left(\boldsymbol{r}\right)} \hspace{-0.2cm} \frac{\rho^{*}\left[k,t\right]}{q}\right]\right]  \\
  =\sum_{\left(k,t\right)\in\mathbb{V}\left(r_{j}\right)-\mathbb{V}\left(r_{j}'\right)}-\frac{\rho^{*}\left[k,t\right]}{m^{\left(k,t\right)}\left(\boldsymbol{r}\right)}. \hspace{4.4cm} 
\end{split} \label{eq:smplersub}
\end{equation}
%

Substituting equations \eqref{eq:smpleradd} and \eqref{eq:smplersub}
into equation \eqref{eq:simplerV} gives us that
\begin{equation}
\begin{split}
\Phi^{V}\left(r_{j}',\boldsymbol{r}_{-j}\right)-\Phi^{V}\left(\boldsymbol{r}\right) \hspace{4cm} \\
  =\left[\sum_{\left(k,t\right)\in\mathbb{V}\left(r_{j}'\right)-\mathbb{V}\left(r_{j}\right)} \hspace{-0.5cm} \frac{\rho^{*}\left[k,t\right]}{m^{\left(k,t\right)}\left(\boldsymbol{r}\right)+1}\right]- \hspace{-0.2cm} \sum_{\left(k,t\right)\in\mathbb{V}\left(r_{j}\right)-\mathbb{V}\left(r_{j}'\right)} \hspace{-0.5cm} \frac{\rho^{*}\left[k,t\right]}{m^{\left(k,t\right)}\left(\boldsymbol{r}\right)} \\
  =U_{j}^{V}\left(r_{j}',\boldsymbol{r}_{-j}\right)-U_{j}^{V}\left(\boldsymbol{r}\right). \hspace{4.4cm} \label{eq:closerV}
\end{split}
\end{equation}
%

Substituting equations \eqref{eq:eqE} and \eqref{eq:closerV} into equation
\eqref{eq:lhs expand}, gives us that
\begin{equation}
\begin{split}
\Phi\left(r_{j}',\boldsymbol{r}_{-j}\right)-\Phi\left(\boldsymbol{r}\right) \hspace{5cm} \\
=\left[U_{j}^{V}\left(r_{j}',\boldsymbol{r}_{-j}\right)-U_{j}^{V}\left(\boldsymbol{r}\right)\right]+\left[U_{j}^{E}\left(r_{j}',\boldsymbol{r}_{-j}\right)-U_{j}^{E}\left(\boldsymbol{r}\right)\right].
\end{split}\label{eq:nearly}
\end{equation}
%

Now equation \eqref{eq:Uadd} can be used to simplify the right hand
side of equation \eqref{eq:nearly}, and this yields the right hand side
of equation \eqref{eq:pot goal}, as required. \hfill \QEDclosed
}

\section{Proof of Theorem \ref{THM:BRPOLY}} \label{app:brpoly}

 The key proof idea is to show that the problem of a generic user $j$ finding a best response
strategy in a strategy profile $\boldsymbol{r}$ can be reformulated as the problem of finding a longest path in a weighted directed acyclic graph $\mathcal{G}^{*}$,
which can be solved in polynomial time.

  To construct such a graph $\mathcal{G}^{*}$, we construct a weighted directed graph $\mathcal{G}=\left(\mathcal{V},\mathcal{E}\right).$ Recall that $\mathbb{V}$ is defined in Definition \ref{z:tt points} and $\mathbb{E}$ is defined in Definition 
  \ref{z:ttp points}. The vertex set $\mathcal{V}=\bigcup_{s\in\mathcal{R}_{j}}\mathbb{V}\left(s\right)$
corresponds to the set of task-time points $(k,t)$, which lie on routes
which are feasible for user $j.$ The directed edge set $\mathcal{E}=\bigcup_{s\in\mathcal{R}_{j}}\mathbb{E}\left(s\right)$
corresponds to the set of all pairs $\left[\left(k,t\right),\left(k',t'\right)\right]\in\mathbb{E}\left(s\right)$
such that user $j$ can move between. Each directed
edge $e=\left[\left(k,t\right),\left(k',t'\right)\right]$ of the
graph $\mathcal{G}$ is associated with an edge weight $w\left(e\right)=-g_{j}(e)=-c_{j}^{l[k],l[k']}.$
Each vertex $\left(k,t\right)$ is associated with
a value 
\begin{equation} \label{newsticker}
  \theta\left(\left(k,t\right)\right)=\frac{\rho^{*}\left[k,t\right]}{\left|\left\{ i\in\mathcal{I}:i\neq j,\left(k,t\right)\in\mathbb{V}\left(r_{i}\right)\right\} \right|+1},
\end{equation}
where $\left|\left\{ i\in\mathcal{I}:i\neq j,\left(k,t\right)\in\mathbb{V}\left(r_{i}\right)\right\} \right|$
equals the number of users other than $j$ which have selected task-time
routes that pass through $(k,t)$ in the strategy profile $\boldsymbol{r}.$ Notice that $\theta\left(\left(k,t\right)\right)$
equals the reward that user $j$ would get for visiting task-time point $(k,t).$

A path in a directed graph is a sequence of vertices, where each vertex
is linked to its successor in the sequence. A best response for user $j$
corresponds to a path in the graph $\mathcal{G}$, which starts at the task-time
point $(k_j^{\text{init}},1),$ and maximizes the sum of the weights of the edges, and the values of the vertices along the path.

Starting from graph $\mathcal{G}$ defined above, we form a new graph $\mathcal{G}^{*}=\left(\mathcal{V}^{*},\mathcal{E}^{*}\right),$ 
by replacing each vertex $(k,t)$ in $\mathcal{G}$ with a pair of new vertices $(k,t,0)$
and $(k,t,1).$ These two new vertices are linked by a new directed edge $\left[(k,t,0),(k,t,1)\right],$ which has an edge weight $w^{*}\left(\left[(k,t,0),(k,t,1)\right]\right)=\theta((k,t))$ as in (\ref{newsticker}), equal to the value of the original vertex $(k,t)$.
 
As a result, graph $\mathcal{G}^{*}$ has a vertex set $$\mathcal{V}^{*}=\left\{ \left(k,t,\gamma\right):(k,t)\in\mathcal{V},\gamma\in\left\{ 0,1\right\} \right\},$$
and an edge set $\mathcal{E}^{*}=\mathcal{E}_{1}^{*}\cup\mathcal{E}_{2}^{*}.$ Here set $$\mathcal{E}_{1}^{*}=\left\{ \left[(k,t,1),(k',t',0)\right]:\left[(k,t),(k',t')\right]\in\mathcal{E}\right\},$$
and each edge $\left[(k,t,1),(k',t',0)\right]\in\mathcal{E}_{1}^{*}$
has a weight
$$w^{*}\left(\left[(k,t,1),(k',t',0)\right]\right)=w\left(\left[(k,t),(k',t')\right]\right),$$
with the right hand side being the weight of the edge originally in graph $\mathcal{G}$. Also set $\mathcal{E}_{2}^{*}=\left\{ \left[(k,t,0),(k,t,1)\right]:(k,t)\in\mathcal{V}\right\},$
and each edge $\left[(k,t,0),(k,t,1)\right]\in\mathcal{E}_{2}^{*}$
has a weight $$w^{*}\left(\left[(k,t,1),(k',t',0)\right]\right)=\theta\left(\left(k,t\right)\right),$$
where the right hand side follows (\ref{newsticker}).

By considering the weights of all edges in $\mathcal{G}^{*}$ as lengths,
one can see that a best response for user $j$ in strategy profile $\boldsymbol{r}$ corresponds
to a longest path in $\mathcal{G}^{*}$ that starts at the vertex
$(k_{j}^{init},t,0)$. Such a longest path can be computed in $\mathcal{O}\left(\left|\mathcal{V}^{*}\right|\times\left|\mathcal{E}^{*}\right|\right)=\mathcal{O}\left(K^{3}T^{3}\right)$
time using the algorithm described in \cite{sedgewick_al11}, because $\mathcal{G}^{*}$
is a directed acyclic graph. Since $\mathcal{G}^{*}$ itself can be
constructed in $\mathcal{O}\left(K^{2} T^{2}\right)$ time based on the previous discussions, the entire process for a user to compute the best response has a complexity of $\mathcal{O}\left(K^{3} T^{3}\right).$ \hfill \QEDclosed 

\section{Proof of Lemma \ref{LEM:SOCIALSURPLUS}} \label{app:socialsurplus}

  We prove the lemma by contradiction. Assume that for any optimal solution $\boldsymbol{r}^*$, we can always find a task $q \in \mathcal{K}$ such that $\sum_{i \in \mathcal{I}} y_i^{q}(r_i^*) > 1$. Let us focus on a particular user $j \in \mathcal{I}$ that is assigned to work on task $q$ (i.e., $y_j^{q}(r_i^*) = 1$). 
	We can define another strategy profile $\boldsymbol{r}$, which is the same as $\boldsymbol{r}^*$ for the task allocation of all users, 
	except that task $q$ is not allocated to user $j$. That is, 
\begin{equation} 
		y_i^k(r_i) =
\begin{cases}
  0, & \mbox{if $k = q$ and $i = j$},\\
  y_i^{k}(r_i^*), & \mbox{otherwise.}
\end{cases}
\end{equation}
  Since the coverage of the tasks under $\boldsymbol{r}$ is the same as $\boldsymbol{r}^*$, where $\sum_{i \in \mathcal{I}} y_i^{k}(r_i) \geq 1$ and $\sum_{i \in \mathcal{I}} y_i^{k}(r_i^*) \geq 1$ for each task $k \in \mathcal{K}$, we have from \eqref{equ:reward} that 
\begin{equation} \label{equ:reward_derive}
	\text{reward}(\boldsymbol{r}) = \text{reward}(\boldsymbol{r}^*).
\end{equation}
  Since user $j$ works on one task less in $r_j$ than in $r_j^*$, from \eqref{equ:triangle}, we have 
\begin{equation} \label{equ:costj_derive}
	\text{cost}_j(r_j) \leq \text{cost}_j(r_j^*).
\end{equation}
  However, the movement costs of other users remain the same such that
\begin{equation} \label{equ:costi_derive}
	\text{cost}_i(r_i) = \text{cost}_i(r_i^*), \, \forall \, i \in \mathcal{I} \backslash \{j\}.
\end{equation}
  Overall, substituting \eqref{equ:reward_derive}, \eqref{equ:costj_derive}, and \eqref{equ:costi_derive} into \eqref{equ:surplus}, we have 
\begin{equation} 
	\text{surplus}(\boldsymbol{r}^*) \leq \text{surplus}(\boldsymbol{r}),
\end{equation}
which means that strategy profile $\boldsymbol{r}$ is also an optimal solution. This leads to a contradiction, and hence proves the theorem. \hfill \QEDclosed 


\opt{opt2}{
\subsection{Discussion of multiple solutions of the social surplus maximization problem} \label{sec:multiplesoln}

  We note that there can be other optimal solutions that do not satisfy Lemma 1. However, these solutions can be obtained from the optimal solution in Lemma 1, when there are no \emph{extra} movement costs for users to work on some tasks, so that there can be more than one user working on a particular task in the optimal solution.
	
	As an example, consider the case with two users (i.e., $I = 2$) and three tasks (i.e., $K = 3$). For the execution time of the tasks, we assume that $t[1] = 2$, $t[2] = 3$, and $t[3] = 4$. For the locations of tasks, we assume that the three tasks are located on a straight line, where the movement cost of user $1$ satisfy the condition
\begin{equation} 
	c_1^{1,2} + c_1^{2,3} = c_1^{1,3}.
\end{equation}
That is, the location $l[2]$ of task $2$ is between the locations $l[1]$ and $l[3]$ of tasks $1$ and $3$.
	
	Assuming that the strategy profile with
	$$r_1 = \bigl((k_1^{\text{init}}, 1), (1,2), (3,4) \bigr)$$ and $$r_2 = \bigl((k_2^{\text{init}}, 1), (2,2), (2,3), (2,4)\bigr)$$
	is an optimal solution of the social surplus maximization problem, which satisfies the condition in Lemma 1.
	From this strategy profile, we can define another strategy profile
		$$\bar{r}_1 = \bigl((k_1^{\text{init}}, 1), (1,2), (2,3), (3,4) \bigr)$$ and $$\bar{r}_2 = \bigl((k_2^{\text{init}}, 1), (2,2), (2,3), (2,4)\bigr),$$
		where both users work on task $2$ at time $3$. Notice that the social surplus achieved under this strategy profile $\boldsymbol{\bar{r}}$ is the same as the strategy profile $\boldsymbol{r}$, since user $1$ does not need to incur any extra movement cost in completing task $2$. 
}

\opt{opt1}{ 
\section{Proof Sketch of Theorem \ref{THM:NP_HARD}} \label{app:NP_hard}
}
\opt{opt2}{ 
\section{Proof of Theorem \ref{THM:NP_HARD}} \label{app:NP_hard}
}
  We prove the NP-hardness by \emph{restriction} \cite{garey_ca79}: We show that finding the social surplus maximization solution in a special case of a TSG can be transformed into a 3-dimensional matching decision problem, which is NP-complete \cite{garey_ca79, kleinberg_ad05}. 
	

\begin{definition} \label{def:3dmatch} 
  \normalfont{(3-dimensional matching) }
  Let $\mathcal{X}$, $\mathcal{Y}$, and $\mathcal{Z}$ be three finite disjoint sets. Let $\mathcal{R} \subseteq \mathcal{X} \times \mathcal{Y} \times \mathcal{Z}$ be a set of ordered triples, i.e., $\mathcal{R} = \{ (x,y,z): x \in \mathcal{X}, y \in \mathcal{Y}, z \in \mathcal{Z} \}$. $\mathcal{R}' \subseteq \mathcal{R}$ is a \emph{3-dimensional matching} if for any two different triples $(x_1, y_1, z_1) \in \mathcal{R}'$ and $(x_2, y_2, z_2) \in \mathcal{R}'$, we have $x_1 \neq x_2$, $y_1 \neq y_2$, and $z_1 \neq z_2$.
\end{definition}

\begin{definition} \label{def:3dmatchproblem} 
  \normalfont{(3-dimensional matching decision problem) }
	Suppose $|\mathcal{X}| = |\mathcal{Y}| = |\mathcal{Z}| = M$. Given an input $\mathcal{R}$ with $|\mathcal{R}| \geq M$, decide whether there exists a 3-dimensional matching $\mathcal{R}' \subseteq \mathcal{R}$ with the maximum size $|\mathcal{R}'| = M$.
\end{definition}

  Consider a restricted TSG, which is a special case of a TSG, which we place the following restrictions:	
	
(a) Tasks and time slots: There are $T = 3$ time slots, and there are $M$ tasks in each time slot (hence a total of $3M$ tasks in $3$ time slots). Sets $\mathcal{X}$, $\mathcal{Y}$, and $\mathcal{Z}$ represent the sets of available tasks in the three time slots, where $|\mathcal{X}| = |\mathcal{Y}| = |\mathcal{Z}| = M$.

(b) Task-time route: Set $\mathcal{R}$ represents the set of available task-time routes of all the users. 
	Assume that each user $i$ can only choose one particular available task-time route $r_i \in \mathcal{R}$.
  We assume that the number of users $I$ is large enough, such that the task-time routes of all the users cover all the tasks in $\mathcal{X} \cup \mathcal{Y} \cup \mathcal{Z}$, so $|\mathcal{R}| = I \geq M$.
	$\mathcal{R}' \subseteq \mathcal{R}$ in Definition \ref{def:3dmatchproblem} represents a feasible task allocation. We assume that a user, whose route is not chosen in $\mathcal{R}'$, will not work on any real task.
		
(c) Large reward: The reward of a task is larger than the movement cost to work on the task, i.e., $\rho[k] > c_i^{l[k'],l[k]}$ for each user $i \in \mathcal{I}$ and all tasks $k,k' \in \mathcal{K}$.

(d) User-independent movement cost: The movement cost only depends on its destination location and is independent of its initial location, such that $c_i^{l,l'} = c[l']$ for all $i \in \mathcal{I}$ and $l,l' \in \mathcal{L}$, where $c[l']$ is the movement cost of destination $l' \in \mathcal{L}$. 

	In the restricted TSG, first, restrictions (c) and (d) imply that we can maximize the social surplus by covering \emph{all} the tasks with \emph{any} available users. 	
  Second, Lemma \ref{LEM:SOCIALSURPLUS}\opt{opt2}{ (and the discussion in Appendix \ref{sec:multiplesoln})} implies that we can focus on an optimal solution, where each task should be allocated to \emph{at most one} user.
 So the optimal task allocation should not contain any overlapping components (i.e., multiple users working on the same task) as defined in Definition \ref{def:3dmatch}. 
	Putting the above discussions together, we know that in the social surplus maximization solution, \emph{every} element of $\mathcal{X} \times \mathcal{Y} \times \mathcal{Z}$ (i.e., every task) should be contained in \emph{exactly one} of the triples (i.e., the task-time routes) in $\mathcal{R}'$. 
	In other words, $\mathcal{R}' \subseteq \mathcal{R}$ is the optimal task allocation. So the surplus maximization problem can be transformed to a 3-dimensional matching decision problem, which is NP-complete \cite{garey_ca79, kleinberg_ad05}. 
	By restriction, we establish that the problem of finding the social surplus maximization solution of the TSG is NP-hard. 
	\opt{opt1}{Full details of the proof can be found in \cite{cheung_dt15_arxiv}.} \hfill \QEDclosed 

\opt{opt2}{
\begin{figure}[t]
 \centering
   \includegraphics[width=8cm, trim = 4cm 3.5cm 4cm 4cm, clip = true]{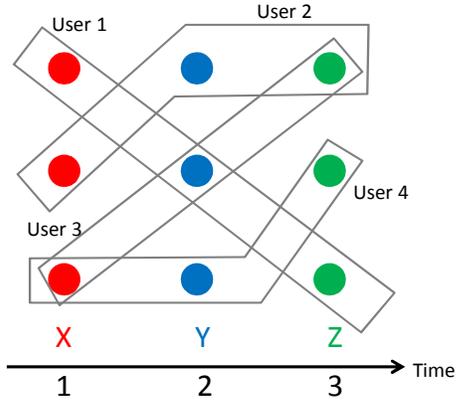} 
 \caption{An example of 3-dimensional matching. Here, the set $\mathcal{R}$ consists of the four grey areas, which represent the routes of users $1$ to $4$. The solution of the 3-dimensional matching problem is the set $\mathcal{R}'$ that consists of the routes of users $1$, $2$, and $4$. It is also the solution of the social surplus maximization problem.}
\label{fig:3dmatch}
\end{figure}
}

\section{Proof of Lemma \ref{LEM:GREEDY}} \label{app:greedy}

  The sorting of execution time in line 6 takes $\mathcal{O}(K \log K)$ time \cite{kleinberg_ad05}. 
  For each task, besides some simple computations, most of the time is consumed for sorting the movement costs of the users in line 9, and it takes a running time of $\mathcal{O}(I \log I)$. Since we generally have more users than tasks, i.e., $I \log I > \log K$, the total run time with $K$ tasks is $\mathcal{O}(KI \log I)$.
\hfill \QEDclosed

\bibliographystyle{abbrv}
\bibliography{mybibfile}

\end{document}